%% file: main.tex
\documentclass{article}

\usepackage{microtype}
\usepackage{graphicx}
\usepackage{subfigure}
\usepackage{booktabs} % for professional tables
\usepackage{multirow}
\usepackage{hyperref}
\usepackage[normalem]{ulem}

\newcommand{\themethod}{DiscoBAX}

\usepackage[accepted]{icml2023}

\usepackage{amsmath}
\usepackage{amssymb}
\usepackage{mathtools}
\usepackage{amsthm}
\usepackage{setspace}
\input{math_commands.tex}
\usepackage[capitalize,noabbrev]{cleveref}

\usepackage[textsize=tiny]{todonotes}

\icmltitlerunning{\themethod{}: Discovery of Optimal Intervention Sets in Genomic Experiment Design}

\begin{document}

\twocolumn[
\icmltitle{\themethod{}: Discovery of Optimal Intervention Sets \\ in Genomic Experiment Design}

\icmlsetsymbol{equal}{*}
\icmlsetsymbol{senior}{$\dagger$}

\begin{icmlauthorlist}
\icmlauthor{Clare Lyle}{equal,oxford,deepmind}
\icmlauthor{Arash Mehrjou}{equal,senior,gsk}
\icmlauthor{Pascal Notin}{equal,oxford} \\
\icmlauthor{Andrew Jesson}{oxford}
\icmlauthor{Stefan Bauer}{helmholtz,tum}
\icmlauthor{Yarin Gal}{oxford}
\icmlauthor{Patrick Schwab}{senior,gsk}
\end{icmlauthorlist}

\icmlaffiliation{deepmind}{Google DeepMind}
\icmlaffiliation{gsk}{GlaxoSmithKline}
\icmlaffiliation{oxford}{University of Oxford}
\icmlaffiliation{helmholtz}{Helmholtz AI}
\icmlaffiliation{tum}{Technical University of Munich}

\icmlcorrespondingauthor{Clare Lyle}{clarelyle@deepmind.com}
\icmlcorrespondingauthor{Arash Mehrjou}{arash@distantvantagepoint.com}
\icmlcorrespondingauthor{Pascal Notin}{pascal.notin@cs.ox.ac.uk}

\icmlkeywords{Machine Learning, ICML, Experiment Design, Active Learning, Genomics}

\vskip 0.3in
]

\printAffiliationsAndNotice{$^*$Equal contribution, $^\dagger$ Senior authorship} % otherwise use the standard text. % otherwise use the standard text.

\begin{abstract}
The discovery of therapeutics to treat genetically-driven pathologies relies on identifying genes involved in the underlying disease mechanisms. Existing approaches search over the billions of potential interventions to maximize the expected influence on the target phenotype. However, to reduce the risk of failure in future stages of trials, practical experiment design aims to find a set of interventions that maximally change a target phenotype via diverse mechanisms. 
We propose DiscoBAX, a sample-efficient method for maximizing the rate of significant discoveries per experiment while simultaneously probing for a wide range of diverse mechanisms during a genomic experiment campaign.
We provide theoretical guarantees of approximate optimality under standard assumptions, and conduct a comprehensive experimental evaluation covering both synthetic as well as real-world experimental design tasks. 
DiscoBAX outperforms existing state-of-the-art methods for experimental design, selecting effective and diverse perturbations in biological systems.
\end{abstract}

\section{Introduction}
\label{sec:introduction}

\begin{figure*}[t!]
    \centering
    \includegraphics[width=14cm]{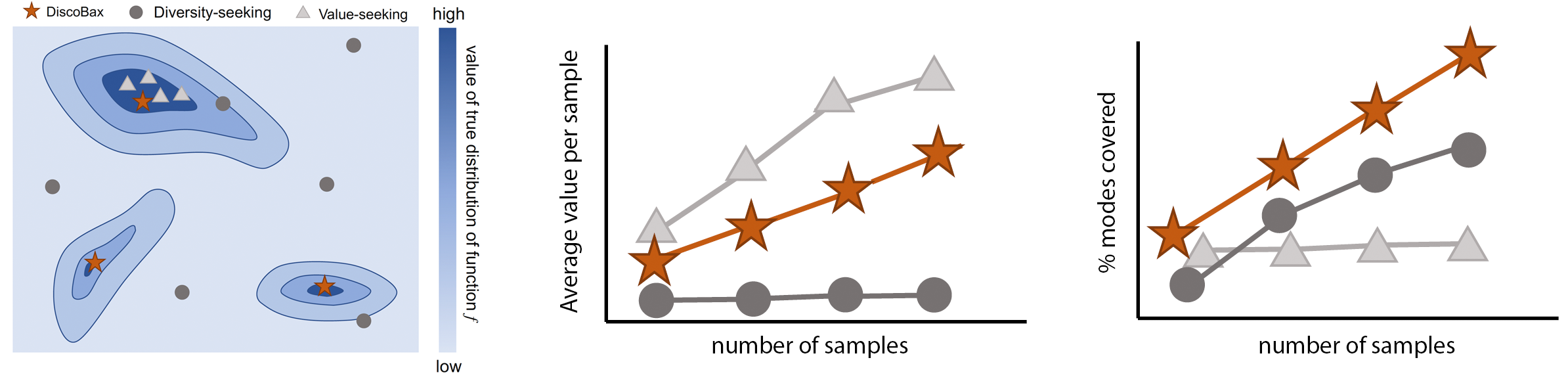}
    
    \caption{We compare \themethod{} (orange star) to existing diversity-seeking (dark grey circle) and value-seeking (light grey triangle) batch active learning policies. \themethod{} aims to recover a maximally diverse set of interventions with values above a pre-defined threshold from a given underlying distribution. This aim contrasts with value-seeking strategies focusing on maximizing value and diversity-seeking strategies focusing on maximizing coverage. We expect \themethod{} to design genomic experiments yielding high value findings that maximize mode coverage. As discussed in \S~\ref{sec:introduction}, the diversity of selected interventions is highly desirable to increase the chances that at least some of these interventions will succeed in subsequent stages of the drug discovery pipeline.}
    \label{fig:overview}
\end{figure*}

Genomic experiments probing the function of genes under realistic cellular conditions are the cornerstone of modern early-stage drug target discovery and validation; moreover, they are used to identify effective modulators of one or more disease-relevant cellular processes. 
These experiments, for example using Clustered Regularly Interspaced Short Palindromic Repeats (CRISPR) \citep{jehuda2018genome} perturbations, are both time and resource-intensive \citep{dickson2004key, dickson2009cost, dimasi2016innovation, berdigaliyev2020overview}. 
Therefore, an exhaustive search of the billions of potential experimental protocols covering all possible experimental conditions, cell states, cell types, and perturbations \citep{trapnell2015defining,hasin2017multi,worzfeld2017unique,chappell2018single,MACLEAN201832,chappell2018single} is infeasible even for the world's largest biomedical research institutes.

To mitigate the chances of failure in subsequent stages of the drug design pipeline, it is desirable for the subset of precursors selected in the target identification stage to operate on diverse underlying biological mechanisms \citep{nica2022evaluating}. 
That way, if a promising candidate based on in-vitro experiments triggers undesirable outcomes when tested in-vivo (e.g., unexpected side effects), other lead precursors relying on different pathways might be suitable replacements that are not subject to the same issues. This two-phase maximization problem diverges from standard formulations of Bayesian optimization or active learning. In particular, the noisy measurements obtained by the experimenter don't correspond to the objective of interest, but are only correlated with this outcome via some unknown mechanism. Thus even in the limit of infinite intermediate phenotype measurements, it is not possible to identify the maximum of the objective function.

Our first contribution is formalizing this problem in order to identify properties of an optimal solution. Mathematically, finding a diverse set of precursors corresponds to identifying and sampling from the different modes of the black-box objective function mapping intervention representations to the corresponding effects on the disease phenotype (\S~\ref{sec:background}).
Existing machine learning methods for iterative experimental design (e.g., active learning, Bayesian optimization) have the potential to aid in efficiently exploring this vast biological intervention space. However, to our knowledge, there is no method geared toward identifying the modes of the underlying black-box objective function to identify candidate interventions that are both effective and diverse (\S~\ref{sec:related-work}).

To this end, we introduce \themethod{} - a sample-efficient Bayesian Algorithm eXecution (BAX) method for discovering genomic intervention sets with both high expected change in the target phenotype and high diversity to maximize chances of success in the following stages of drug development (\Cref{fig:overview}), which we formalize as set-valued maximization problem (Equation~\ref{eq:master_equation}). 
After providing theoretical guarantees on the approximate optimality of the presented approach under standard conditions, we perform a comprehensive experimental evaluation in both synthetic and real-world datasets. These experiments show that \themethod{} outperforms existing state-of-the-art active learning and Bayesian optimization methods in designing genomic experiments that maximize the yield of findings that could lead to the discovery of new potentially treatable disease mechanisms. The implementation of \themethod{} and the code to reproduce the experimental results are publicly available in \url{https://github.com/amehrjou/DiscoBAX}.

Our contributions are as follows:

\begin{itemize}
\item We give a formalization of the gene target identification problem (\S~\ref{sec:problem-setting}) and discuss limitations of existing methods in addressing this problem (\S~\ref{sec:related-work}).

\item We develop \themethod{} - a sample-efficient BAX method for maximizing the rate of significant discoveries per experiment while simultaneously probing for a wide range of diverse mechanisms during a genomic experiment campaign (\S~\ref{sec:method}).

\item Leveraging insights from the mathematical structure of our formalization, we provide theoretical guarantees that substantiate the optimality properties of \themethod{} (\S~\ref{sec:method} and Appendix~\ref{appendix:submodular}).

\item We conduct a comprehensive experimental evaluation covering both synthetic as well as real-world experimental design tasks that demonstrate that \themethod{} outperforms existing state-of-the-art methods for experimental design in this setting (\S~\ref{sec:experiments}).
\end{itemize}

%%%%%%%%%%%%%%%%%%%%%%%%%%

\section{Background and Notation}
\label{sec:background}

Genomic experimentation is an early stage in drug discovery where geneticists assess the effect of genomic interventions on moving a set of disease-relevant phenotypes to determine suitable drug targets.

To formalize this process, we assume a black-box function, $f:\gG \to \mathbb{R}$, that maps each gene, $\g \in \gG$, to the value, $f(\g)$, corresponding to the magnitude of phenotypic change under gene knock out.
The set, $\gG$, is finite, $|\gG|=m<\infty$, because there are a limited number of protein-encoding genes in the human genome ($\approx 20,000$) \citep{pertea2018chess}, and can be represented by either the set of integers or one-hot vectors with dimension $m$.
However, biologically informed embeddings, $\X: \gG \to \Xcal$, are often preferred to represent genes for their potential to capture genetic, functional relationships. We assume that gene embeddings, $\X(g) = \x\in\Xcal \subseteq \mathbb{R}^d$, are sets of $d$-dimensional real vectors, with $m$ distinct members, $|\Xcal| = m$, thus, we use $f(\g)$ and $f(\x)$ interchangeably, where $\x$ is the embedding of the gene $\g$.

In drug development, a candidate target must meet several criteria to proceed to subsequent stages in the development pipeline. 
For example, engaging the target -- down- or up-regulating the gene -- must move the phenotype \emph{significantly} in the desired direction. 
Such genes are called ``top-movers" of the phenotype. 
We can define the $K$ top-movers for a given phenotype as members of the set, $\Xcal=\{\x_1, \x_2, \ldots, \x_m\}$, corresponding to the $K$ largest values of $\{f(\x_1), f(\x_2), \ldots, f(\x_m)\}$.
However, each evaluation of the phenotype change, $f$, requires a CRISPR-Cas9 knockout experiment in the lab, which makes exhaustive experimentation infeasible even for the most resourceful institutions. 
Hence in practice, the experimentation budget is limited to $T \ll m$ experiments. 
Instead of choosing the $K$ top-movers (requiring phenotype change knowledge, $f(\x)$, for all inputs $\x \in \Xcal$), a more practical approach is to form the subset, $\Xcal_c \subseteq \Xcal$, of genes that when knocked out lead to a change in the phenotype, $f(\x)$, larger than a selected threshold value, $c$, i.e. $\Xcal_c \coloneqq \{\x\in \Xcal : f(\x) \geq c \}.$

A critical aspect distinguishing the drug discovery pipeline from the standard Bayesian Optimization setting is that we do not seek to identify a single point which maximizes the unknown function $f$; rather, we wish to identify a computable property of $f$ using a limited number of evaluations of $f$. To do so we will leverage Bayesian Algorithm Execution (BAX), proposed by \citet{neiswanger2021bayesian}, which is designed precisely to identify the output, $\Out_\gA \coloneqq \Out_{\gA}(f)$, of an algorithm, $\gA$, run on a function, $f$, by evaluating the function on a budgeted set of inputs, $\{\x_i\}_{i=1}^T \in \gX$. 
Estimating a computable property, i.e. the output of the algorithm $\gA$, is done by positing a probabilistic model for $f$ for estimating $\Out_\gA$.
Data is acquired by searching for the value $\x \in \gX$ that maximizes the mutual information, $I(\mathrm{Y}_\x; \Out_{\gA} \mid \gD_t)$, between the function output, $\mathrm{Y}_{\x}$, and the algorithm output, $\Out_{\gA}$. 
BAX assumes that functional output instances, $\y_{\x}$, of the function, $f$, can be observed for each acquired $\x$.
The acquisition of data is sequential, where the information gain maximization procedure leads to a dataset of observations,  $\gD_t \coloneqq \{(\x_i, \y_{\x_i})\}_{i=1}^{t-1}$, at step $t \in [T]$.
BAX can be used in conjunction with a number of algorithms, such as determining the superlevel set (i.e. $\Xcal_c$), computing integrals, or finding local optima of $f$. 
Given that genomic experimentation seeks to find a diverse set of genes corresponding to the modes of $f$, the BAX framework is well suited to our task.

Concretely, BAX acquisition functions select points by maximizing the expected information gain (EIG) obtained from each point about the output of the algorithm. Crucial to the applicability of BAX to our problem setting is the tractability of accurate approximators of the EIG for algorithms which, like the one we will propose, return a subset of their inputs. The exact computation of the EIG for arbitrary algorithms is not generally tractable; however, \citet{neiswanger2021bayesian} present an approximation that only requires the computation of the entropy of the distribution over function values conditioned on algorithm outputs.
\begin{equation}\label{eq:eigv}
\begin{split}
     \mathrm{EIG}^v_t(\x, \gD_t) =  H&(\ip(\x) | \gD_t) - \\ &\mathbb{E}_{p(S | \gD_t)}[ H(\ip(\x) | S, \gD_t) ].   
\end{split}
\end{equation}
When the model $P$ is a Gaussian Process (GP), both quantities are straightforward to compute: the first is the entropy of the GP's predictive distribution at $\x$, and we can estimate the second by conditioning a posterior on the values of elements in the set $S$. 
Monte Carlo approximation of this quantity is possible when the model $P$ does not permit a closed form.
%%%%%%%%%%%%%%%%%%%%%%%%
\section{Problem Setting}
\label{sec:problem-setting}

A primary challenge in the drug discovery pipeline is the discrepancy in outcomes between \textit{in vitro} experimental measurements and \textit{in vivo} outcomes.
Where \textit{in vitro} experimental data can quantify the effect of a gene knockout on a specific aspect of a cellular phenotype in a petri dish, \textit{in vivo} interactions between the drug and the organism may lead to weaker effect sizes or toxicity. 
The drug discovery pipeline consists of stages, starting by testing a set of candidate interventions and then proceeding by selecting a subset of promising candidates to pass on for further development. 
For example, one might test a broad range of gene knockouts on cell cultures and then select a subset of promising gene candidates to evaluate in animal models. 
These trials can be expensive, so it is desirable to weed out potentially ineffective or toxic candidates before this phase. 
To do so, researchers can leverage heuristic score functions that predict the "drug-likeness" or likelihood of toxicity of a compound \citep{jimenez2020drug}. 
Considering a diverse set of candidate interventions, where each intervention applies to a different mechanism in the disease phenotype, is also of use as it increases the likelihood of at least one candidate succeeding in the subsequent phase.

We formalize this setting as an optimization problem in which the optimizer has access to a measurement which is correlated with the outcome of interest; however, some assumed noise model distorts this quantity before yielding the primary objective function value. 
We formalize our search space (i.e., the set of available genes, though in principle this could be any set) $\gG=\{\g_1, \dots, \g_m \}$, for which we have some phenotype measurement $\ip$. 
We will primarily refer to $\ip$ as a function from \textit{features} to phenotype changes, but it is equivalent to expressing $\ip$ as a function on genes $\gG$. 
The subscript `$\textrm{ip}$' stands for \emph{intermediate phenotype} as it is not the actual clinical measurement caused by the gene knockout. 
Instead, it is a measurement known to correlate with a disease pathology and is tractable in the lab setting (see \Cref{sec:biology_background} for detailed formalization). 
In this paper, we will assume the phenotype change is a real number $\ip(\x) \in \mathbb{R}$; however, given suitable modeling assumptions, it is possible to extend our approach to vector-valued phenotype readouts. 
We also define a function called the \emph{disease outcome}, $\dis$, which is composed of $\ip$ and factors outside the biological pathway, such as toxicity of a molecule that engages with a target gene. 
The noise component, $\eta$, encapsulates all these extra factors.

In practice, $\eta$ will depend on the nature of the biological systems under consideration, and could take on a variety of forms of varying degrees of complexity. The only assumption we make is that $\eta$ is a locally smooth function of x. From a biological standpoint, this implies that the noise $\eta$ for two interventions on similar genes will be similar (e.g., if one leads to toxicity, the other one will likely do so as well). Here, we illustrate two tractable formulations of the relationship between the disease outcome, $\dis$, and the \textit{in vitro} phenotype, $\ip$.

\begin{enumerate}
    \item \textbf{Multiplicative Bernoulli noise:}
    \begin{equation}
        \dis(\x; \eta) = \ip(\x) \eta(\x)
    \end{equation}
    where $\eta(\x) \in \{0,1\}, \forall \x \in \gG$, and $\eta$ is sampled from a Gaussian process classification model. This setting presents a simplified model of drug toxicity: $\eta$ corresponds to a binary indicator of whether or not the drug is revealed to exhibit unwanted side effects in future trials. The multiplicative noise model assumes that the downstream performance of an intervention is monotone with respect to its effect on the phenotype, conditional on the compound not exhibiting toxicity in future trials.
    In our experiments, we assume $\eta$ exhibits correlation structure over inputs corresponding to a GP classification model, and construct the kernel $K_{\Xcal}$ of this GP to depend on some notion of distance in the embedding space $\Xcal$. 
    \item \textbf{Additive Gaussian noise:}
    \begin{equation} \label{eq:objective}
        \dis(\x; \eta) = \ip(\x) + \eta(\x) \quad \eta \sim \textrm{GP}(\mathbf{0}, K_{\Xcal})
    \end{equation}
    where $\eta:\gG \rightarrow \RR$ is drawn from a Gaussian process model with kernel $K_{\Xcal}$. 
    In this case, we assume that the unforeseen effects of the input $\x$ are sufficiently numerous to resemble a Gaussian perturbation of the measured in vitro phenotype $\ip(\x)$. 
\end{enumerate}

Notice that in the above models, noise is an umbrella term for everything that affects the fitness of a target but is not part of the biological pathway from the gene to the phenotype change. Therefore, the choice of noise distribution and how it affects the outcome is a modelling assumption that is intended to capture coarse inductive biases known to the researcher.
We additionally seek out a \textit{set} of interventions $\sg \subset \gG$ of some fixed size $|\sg|=k$ whose elements cause the maximum expected change (for some noise distribution) in the disease outcome. 
In other words, we seek an intervention that best moves the disease phenotype, which will be the best candidate drug. This goal is distinct from either sampling the super-level-sets of $\ip$ or finding the set $\sg$ with the best average performance. 
Instead, we explicitly seek to identify a set of points whose toxicity or unintended side effects will be minimally correlated, maximizing the odds that at least one will succeed in the subsequent trials. 
We thus obtain a set-valued maximization problem
\begin{equation}
\label{eq:master_equation}
   \max_{S \subseteq \gX} \mathbb{E}_{\eta}\bigg [ \max_{\x \in S} \dis(\x; \eta) \bigg ] \; .
\end{equation}

This compact formula is critical to attain our overarching objective: identifying interventions with both a large impact on the phenotype of interest and with high diversity to increase the chance of success of some of them in the subsequent steps of the drug discovery pipeline. An illustrative example is provided in \Cref{fig:master_equation_intuition} in Appendix~\ref{apx:deeper-insights} to provide further intuition into this formula.

The general formulation of this problem is NP-hard \citep{goel2010how}; therefore, we propose a tractable algorithm that provides a constant-factor approximation of the optimal solution by leveraging the submodular structure of the objective under suitable modeling assumptions~\cite{Golovin2010AdaptiveST}. 
Given such an algorithm, our task is the active learning problem of optimally querying the function, $\ip$, given a limited number of trials, $T$, to accurately estimate the algorithm's output on the ground-truth dataset. 

Importantly, this formulation allows us to decouple modeling the measured phenotype, $\ip$, from modeling the noise $\eta$. 
For example, we might make the modeling assumption that we sample $\ip$ from a GP with some kernel $k_1$ and that $\eta$ is a Bernoulli random variable indicating the safety of the compound.

%%%%%%%%%%%%%%%%%%%%%%%%
\section{Method}
\label{sec:method}

Various methods exist for efficiently optimizing black-box functions; however, our problem setting violates several assumptions underlying these approaches. 
In particular, while we assume access to intermediate readouts $\ip$, the actual optimization target of interest $\dis$ is not observable.
Further, we seek to find a \textit{set} of interventions that maximize its expected value under some modeling assumptions. 
These two properties render a broad range of prior art inapplicable.
Active sampling methods do not prioritize high-value regions of the input space. 
Bayesian optimization methods assume access to the ground-truth function outputs (or a noisy observation thereof). 
And Bayesian algorithm execution approaches based on level-set sampling may not sufficiently decorrelate the hidden noise in the outcome. 

We propose an intervention set selection algorithm in a Bayesian algorithm execution procedure that leverages the modeling assumptions characterized in the previous section.
This method, Subset Discovery via Bayesian Algorithm Execution (DiscoBAX), consists of two distinct parts. 
(1) a subset-selection algorithm obtaining a $1-1/e$-factor approximation of the set that maximizes ~\eqref{eq:objective}, and (2) an outer BAX loop that queries the phenotype readings to maximize the information gain about the output of this algorithm. 
In Section~\ref{sec:algorithm}, we present the idealized form of DiscoBAX and show that it attains an approximately optimal solution. 
Our approach is easily adaptable to incorporate approximate posterior sampling methods, enabling its use with deep neural networks on high-dimensional datasets. 
We outline this practical implementation in Section~\ref{sec:implementation}.

%%%%%%%%%%%%%%%%%%%%%%%%
\subsection{Algorithm}
\label{sec:algorithm}

\textbf{Subset maximization:} we first address the problem of identifying a subset $S \subset \gX$ such that $|S| = k$ which maximizes the value ${\mathbb{E}_{\eta}[ \max_{\x \in S}\dis(\x; \eta)]} $ 
As mentioned previously, the exact maximization of this objective is intractable. To construct a tractable approximation, we propose a submodular surrogate objective, under which the value of an intervention is lower-bounded by zero $\dis^*(\x;\eta)=\max(\dis(\x;\eta), 0)$. 
This choice is motivated by the intuition that any intervention with a negative expected value on the phenotype is equally useless as it will not be considered in later experiment iterations, and so we do not need to distinguish between harmful interventions. 
The resulting function $f(S) = {\mathbb{E}_{\eta}[ \max_{\x \in S}\dis^*(\x; \eta)]} $ will be submodular, and thus Algorithm~\ref{alg:subset-select}, the greedy algorithm, will provide a $1-1/e$ approximation of the optimal solution \citep{nemhauser1978analysis}.

\begin{observation}
    \label{obs:submodularity_of_subsetselection}
    The score function $f: \gP(\gG) \rightarrow \mathbb{R}$ defined by
    \begin{equation}
        f(S) = \mathbb{E}_{\eta}\bigg [\max_{\x \in S} \bigg (\max (0, \dis(\x; \eta)\bigg) \bigg ]
    \end{equation}
    is non-negative, monotone, and submodular.
\end{observation}

We provide proof of this result in Appendix~\ref{appendix:submodular}. In practice, we can estimate the expected value in this objective using Monte Carlo (MC) samples over the noise distribution $\eta$.
Where MC sampling is too expensive, a heuristic that uses a threshold to remove points whose values under $\eta$ are too highly correlated can also obtain comparable results with a reduced computational burden.

\begin{algorithm}[H]
\begin{spacing}{1.1}
             \caption{SubsetSelect }\label{alg:subset-select}
             \begin{algorithmic}
                 \REQUIRE integer $k > 0$, set $\gX$, noise distribution $P(\eta)$, sampled  readouts $\iphat : \gX \rightarrow \mathbb{R}$
                 \STATE $\sg \gets \emptyset$
                \IF {multiplicative noise}
                 \STATE $\dishat(\x; \eta) := \iphat(\x)\eta(\x)$
                 \ENDIF
                \IF {additive noise}
                 \STATE $\dishat(\x; \eta) := \iphat(\x) + \eta(\x)$
                 \ENDIF 
                 \FOR {$i < k$}
                \vspace{-20pt}
                    \STATE \[\sg \gets \sg \cup \{ \argmax_{x \in \gX \setminus S} \mathbb{E}_{\eta}[\max_{y \in S \cup \{x\}} \dishat(\x; \eta)]\}\]
                \vspace{-20pt}
                 \ENDFOR
                 \OUTPUT $\sg$
             \end{algorithmic}
\end{spacing}
\end{algorithm}

\begin{algorithm}[H]
             \caption{DiscoBAX}\label{alg:active-learning}
             \begin{algorithmic}
                 \REQUIRE finite sample set $\gX$, budget $T$, Monte Carlo parameter $\ell \in \mathbb{N}$
                 \STATE $\gD \gets \emptyset$
                 \FOR {$i < T$}
                     \STATE sample $\{\iphat\}_{j=1}^\ell \sim P(\ip | \gD)$ 
                     \STATE $\sg_j \gets$ $\mathrm{SubsetSelect}(\widehat{f_{\text{ip},j}}), \forall j = 1, \dots, \ell$
                     \STATE $\x_i \gets \argmax_{\x\in\gX} \mathrm{EIG}^v(\x, S_{j=1}^\ell)$
                     \STATE query $\ip(\x_i)$
                     \STATE $\gD = \gD \cup \{(\x_i, \ip(\x_i)\}$
                 \ENDFOR
                 \OUTPUT $\gD$
             \end{algorithmic}
             \label{alg:discobax}
\end{algorithm}

\textbf{Active sampling:} because we do not assume prior knowledge of the phenotype function $\ip$, we require a means of selecting potential interventions for querying its value at a specified input $\x$. In practice, running these experiments may incur a cost, and so it is desirable to minimize the number of queries necessary to obtain an accurate estimate of the optimal intervention set. 
BAX \citep{neiswanger2021bayesian} presents an effective active sampling approach to approximate the output of an algorithm using a minimal number of queries to the dataset of interest. 
In our setting, this allows us to approximate the output of \Cref{alg:subset-select} over the set $(\gX, \ip(\gX))$ without incurring the cost of evaluating the effect of every knockout intervention in $\gG$. 
Concretely, this procedure takes as input some probabilistic model $P$ which defines a distribution over phenotype readings $\ip$ conditioned on the data $\gD_t$ seen so far and from which it is possible to draw samples. We consider two noise models in Algorithm~\ref{alg:subset-select}, but the algorithm can be extended to arbitrary noise models by setting $\dishat(x; \eta)$ to be a suitable function of the noise $\eta$.

\textit{A remark on the efficiency of subset maximization \& active sampling. } We emphasize that subset selection is a function called within each active sampling cycle. Hence, the above observation about submodularity refers specifically to \Cref{alg:subset-select} rather than its incorporation in \Cref{alg:active-learning}; without access to the ground truth value of $f$, it is not possible to provide deterministic guarantees on the output of the algorithm. If sample efficiency is not a concern, \Cref{alg:subset-select} can be run on the set of all inputs to obtain the $(1 - 1/e)$-optimal solution. 

We outline this procedure in \Cref{alg:discobax}, and refer to Section~\ref{sec:background} for additional details. 
In the batch acquisition setting, we form batches of size $B$ at each cycle by selecting the $B$ points with the highest $EIG$ values.

\subsection{Practical implementation in high dimensions}
\label{sec:implementation}
When working with high-dimensional input features, we typically leverage Bayesian Neural Networks in lieu of Gaussian Processes. We sample from the parameter distribution via Monte Carlo dropout (MCD) \citep{gal2016dropout}, and rely on Monte Carlo simulation to estimate the quantities introduced in Algorithm~\ref{alg:active-learning}. In particular, the entropy of the posterior distribution is obtained as follows:
    
\begin{equation}
    \begin{split}
    \label{eq:entropy_bnn}
    &H(\y|\x,\gD) = \mathbb{E}_{p(\y|\x, \gD)} \left[-\log p(\y|\x, \gD)\right]\\
                &\approx -\frac{1}{M} \sum_{i=1}^{M} \log p(\y_i| \x, \gD) \\
                &= -\frac{1}{M} \sum_{i=1}^{M} \log \int p(\y_i| f, \x, \gD)p(f|\x, \gD) df\\
                &\approx -\frac{1}{M} \sum_{i=1}^{M} \log \left[\frac{1}{N} \sum_{j=1}^{N} p(\y_i|f_j, \x, \gD) \right]\\
    \end{split}
\end{equation}
where the samples $\{f_j\}_{j=1}^N$ are obtained by sampling from the distribution over model parameters with MCD, and we draw and re-use the same Monte-Carlo samples for both the inner and outer sum. We note that such a nested Monte-Carlo estimator is a biased estimator~\cite{Rainforth2017OnNM}. 
In practice, we use the following approximation derived in \citet{gal2016dropout} (Eq. 8): 
\begin{equation}
\begin{split}
\log p(\y|\x, \gD) &\approx \text{logsumexp}(-\frac{1}{2}\tau\|\y-\y_i\|^{2}) - \log M \\
&- \frac{1}{2}\log 2\pi - \frac{1}{2}\log\tau^{-1}
\end{split}
\end{equation}
with a logsumexp of M terms, $\y_i$ stochastic forward passes through the network, and $\tau$ a precision parameter.

\textit{Remarks on optimality--} Notice that the theoretical guarantees of optimality are provided for the inner loop (\cref{alg:subset-select}) of \themethod{}. The sample-efficiency of the entire algorithm is supported by empirical evidence from a wide range of synthetic and real-world experiments. For further empirical analysis of the sample-efficiency of the BAX procedure, we refer to \citet{neiswanger2021bayesian}.

%%%%%%%%%%%%%%%%%%%%%%%%
\section{Experiments}
\label{sec:experiments}

In the experimental evaluation of \themethod{}, we specifically seek to answer the following questions: 1) Does \themethod{} allow us to reach a better trade-off between recovery of the top interventions and their diversity (\Cref{table:genedisco_results,table:genedisco_detailed_results_interferon,table:genedisco_detailed_results_interleukin,table:genedisco_detailed_results_sars,table:genedisco_detailed_results_leukemia,table:genedisco_detailed_results_tau})? 2) Is the method sample-efficient, i.e., identifies global optima in fewer experiments relative to random sampling or naive optimization baselines (Figure \ref{fig:genedisco_main_results} and \ref{fig:genedisco_main_results_appendix})? 3) Is the performance of \themethod{} sensitive to various hyperparameter choices (Appendix \ref{appendix:genedisco_hypers})? To address these questions, we first focus on experiments involving synthetic datasets (\S~\ref{sec:synthetic}) in which we know the underlying ground truth objective function. We then conduct experiments across several large-scale experimental assays from the GeneDisco benchmark \cite{mehrjou2021genedisco} that cover a diverse set of disease phenotypes.

\subsection{Synthetic Dataset}
\label{sec:synthetic}

\begin{figure*}[t!]
    \centering
    \includegraphics[width=\linewidth]{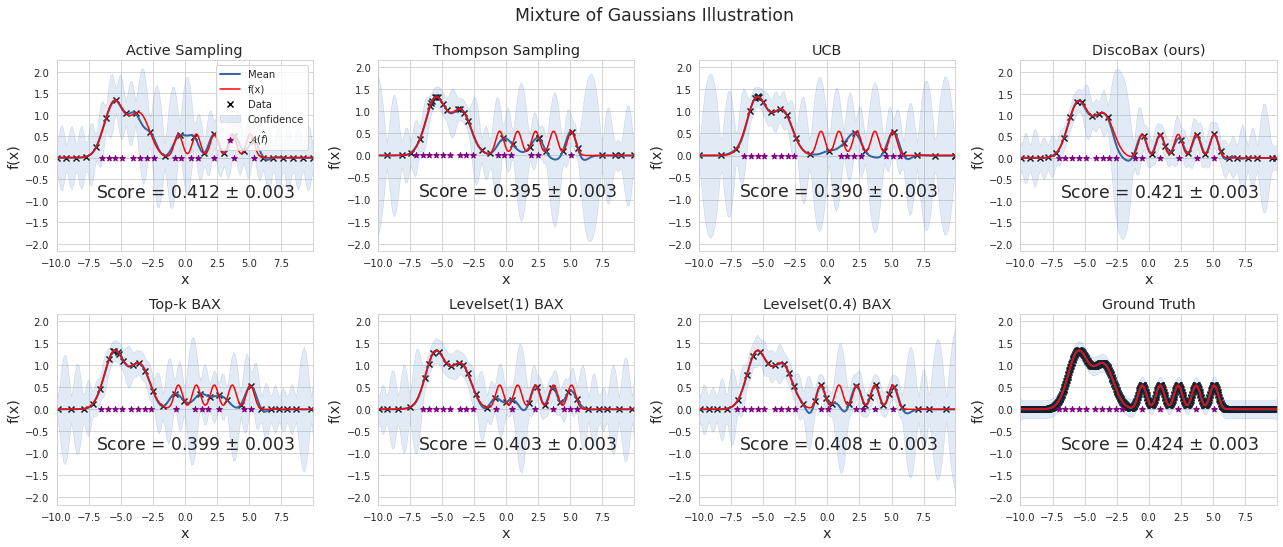}
    % \vspace{-1em}
    \caption{Illustration of failure modes of benchmark acquisition functions in our problem setting: existing methods struggle to accurately capture both the high- and low-valued local optima. We use a batch size equal to one for all methods.}
    \label{fig:toy-example}
\end{figure*}

We begin with a concrete example to illustrate the distinction between the behavior \themethod{} and existing methods. 
The dataset we consider is a one-dimensional regression task on a mixture-of-Gaussians density function $f_{\mathrm{mog}}$. We construct $f_{\mathrm{mog}}$ such that it exhibits several local optima at a variety of values, necessitating a careful trade-off between exploration and exploitation to optimize the DiscoBAX objective. Crucially, exploitation in this setting requires not only an accurate estimation of the global optimum but also an accurate estimation of the local optima. We provide evaluations on additional datasets in Appendix~\ref{apx:synthetic-experiments}.
We consider the following baseline acquisition functions which select the optimal point $\x^*$ to query at each iteration, letting $\mu(\x)$ denote the posterior mean over $\ip(\x)$ and $\sigma^2(\x)$ its variance. We evaluate random sampling, a UCB-like acquisition function, BAX on super-level set and top-k algorithms, Thompson sampling, and uncertainty maximization baselines. Full details are provided in Appendix~\ref{apx:synthetic-experiments}.

In Figure~\ref{fig:toy-example}, we visualize the solutions found by each approach after 30 iterations. We further evaluate the score of each method, computed as ${\mathbb{E}_{\eta} \max_{\x \in S} \ip(\x) \eta(\x) } $, where $\eta$ is drawn from a Bernoulli distribution whose logits are determined by an affine transformation of a sample from a GP with zero mean and radial basis function covariance kernel. This construction ensures a high correlation between the values of nearby inputs and reward sets $S$ whose elements are distant from each other. To select $S$, we use the learned posterior mean $\mu$ from each acquisition strategy as input to Algorithm~\ref{alg:subset-select} and set $S$ to be equal to its output. We observe that most baselines over-exploit the high-value local optima, leading to inaccuracies on the lower optima: Algorithm~\ref{alg:subset-select} is unable to select the optimal subset elements from the lower-value modes and the model score suffers. The advantage of \themethod{} is better visible by looking at the 5 minor modes on the right side of the function. \themethod{} has distributed its experimental budget almost evenly among those modes and has discovered all of them while the other methods either miss some of the modes (no violet star under a mode) or waste extra budget on some modes (more than one violet star under a mode.)

\subsection{GeneDisco Datasets}
\label{sec:genedisco}

\textbf{Datasets \& baselines.} The GeneDisco benchmark \citep{mehrjou2021genedisco} is comprised of five large-scale genome-wide CRISPR assays and compares the relative strengths of nine active learning algorithms (eg., Margin sampling, Coreset) for optimal experimental design. The objective of the different methods is to select the set of interventions (ie., genetic knockouts) with the largest impact on the corresponding disease phenotype.
We include all baselines from the GeneDisco benchmark, as well as six additional approaches: Upper Confidence Bound (UCB), Thompson sampling, JEPIG~\citep{Kirsch2021TestDA}, Top-K BAX and Levelset BAX~\cite{neiswanger2021bayesian}, and \themethod{}.

\textbf{Metrics \& approach. } We define the set of optimal interventions as those in the top percentile of the experimentally-measured phenotype (referred to as `Top-K interventions'). We use Top-K recall to assess the ability of a method to identify the best interventions. To quantify the diversity across the set of optimal interventions, we first cluster interventions in a lower-dimensional subspace (details provided in Appendix~\ref{appendix:clustering}). We then measure the proportion of clusters that are recalled (i.e., any of its members are selected) by a given algorithm over the different acquisition cycles. The geometric mean between Top-K recall and the diversity metric defines the overall score for a method. For all methods and datasets, we perform 25 consecutive batch acquisition cycles (with batch size 32). All experiments are repeated 20 times with different random seeds.

\begin{figure*}[t!]
    \centering
    \includegraphics[width=1.0\linewidth]{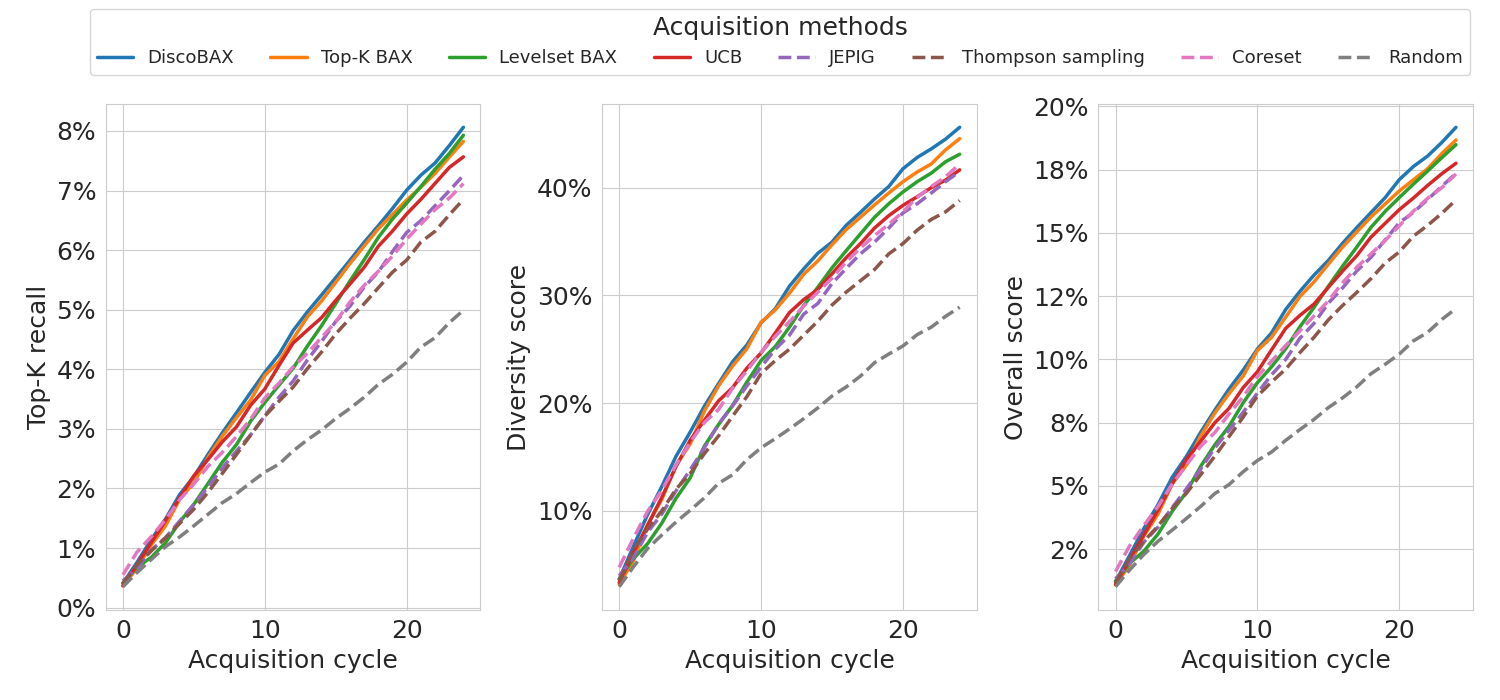}
    \caption{\textbf{Top-K recall, Diversity score and Overall score Vs acquisition cycles across all GeneDisco assays}.}
    \label{fig:genedisco_main_results}
\vspace{-4mm}
\end{figure*}

\textbf{Results \& discussion. } We observe that, across the different datasets, \themethod{} reaches the highest aggregate performance relative to other baselines, as it both recalls a high share of optimal interventions and identifies a diverse set of optimal interventions  (Table~\ref{table:genedisco_results}). It does so in a sample-efficient manner as it achieves high diversity and recall throughout the different acquisition cycles (Fig.\ref{fig:genedisco_main_results}). Note that sample-efficiency is an empirical observation here, not a theoretical property of the algorithm, since it is possible to construct adversarial datasets where a BAX method will attain no better performance than random sampling. 
Looking at the assay-level performance (Appendix~\ref{appendix:genedisco_results}), we observe that the best method varies across assays. Certain methods, such as Coreset or UCB, achieve very high performance on 1 or 2 assays but perform rather poorly in other settings. Compared with Coreset and other active learning approaches, JEPIG performs reliably well across assays. BAX-based approaches, and \themethod{} in particular, tend to perform consistently high across assays. 
Crucially, we find that the performance of \themethod{} is relatively insensitive to the choice of hyperparameters (Appendix~\ref{appendix:genedisco_hypers}), unlike other BAX approaches. Lastly, we note that when the input feature space (ie., the intervention representation) does not correlate much with the disease phenotype of interest, the model being learned tends to perform poorly and we observe no lift between the different methods and random sampling (eg., the SARS-CoV-2 assay from \cite{zhu2021genome}).

\input{genedisco_results.tex}

%%%%%%%%%%%%%%%%%%%%%%%%
\section{Related work}
\label{sec:related-work}

Prior works have studied the application of genomic discovery and method development for diverse target generation. While sharing a philosophical connection to our contribution, the mathematical formalization of the problems these methods seek to solve exhibit subtle but important distinctions from the formulation presented in this paper.

\textbf{Bayesian optimization} 
Bayesian optimization (BO) is concerned with finding the global optimum of a function with the fewest number of function evaluations \citep{snoek2012practical, shahriari2015taking}. 
Since this target function is often expensive-to-evaluate, one typically uses a Gaussian process as a surrogate function \citep{srinivasgaussian}. 
The candidates for function evaluation are then determined through a so-called acquisition function, which is often expressed as the expected utility over the surrogate model. 
Typical utility functions include the expected improvement \citep[EI]{movckus1975bayesian} and probability of improvement \citep[PI]{kushner1964new}. Recent work includes variational approaches \citet{song2022likelihood} which yield a tractable acquisition function whose limiting behavior is equivalent to PI. 
Bayesian optimization has been applied to biological problem settings such as small molecule optimization \citep{griffiths2017constrained,korovina2020chembo,notin_improving_2021} or protein design \citep{moss_boss_2020}. 
While BO bears some resemblance to our problem formulation, these methods cannot naively be applied to our setting as the noisy function observations we receive are not noisy samples of the objective function we seek to maximize.

\textbf{Active learning} 
Active learning approaches focus on prioritizing data points to be labeled based on their informativeness, often with the goal of reaching faster model convergence or reducing annotation costs. 
Different strategies have been proposed to characterize `informativeness'. Diversity-based methods, such as Coreset~\cite{sener2017active} and K-means-based methods~\cite{mehrjou2021genedisco}, emphasize selecting a diverse and representative set of samples. Uncertainty-based methods, such as Top Uncertainty, query the most uncertain points. Bayesian Active Learning by Disagreement (BALD)~\citep{houlsby2011bayesian, kirsch2019batchbald} selects uncertain points that maximize the mutual information between model predictions and parameters, i.e., with the highest Expected Information Gain (EIG) about model parameters.
Since the BALD objective does not take into account the distribution of test points, it may select points with limited relevance. This recently gave rise to methods seeking to instead maximize the EIG about possible future predictions~\cite{Kirsch2021TestDA,Smith2023PredictionOrientedBA}.

\textbf{Bandits} The upper confidence bounds seen in BO originate in the bandit setting \citep{lai1985asymptotically}, in which one can extend the widely-used UCB algorithm to Gaussian processes \citep{ grunewalder2010regret, srinivasgaussian}. 
While both bandits and BO seek to find the maximum of a function, the two problem settings leverage different notions of optimality. BO seeks to \textit{identify} the argmax, whereas bandits seek to \textit{minimize} the number of sub-optimal queries.
Related to bandits and BO, some efforts are made to formulate active learning as a reinforcement learning problem \citep{slade2022deep, Casanova2020Reinforced, konyushkova2017learning, pang2018meta}.
As with BO, the assumption that the learner has access to noisy samples of the objective function is baked into bandit algorithms, making these approaches also unsuitable for our setting.

\textbf{Optimal experiment design} (OED) is a broad umbrella whose scope includes Bayesian Optimization: rather than simply maximizing a parametric function, the task is to adaptively identify an optimal set of experiments to efficiently reach some goal \citep{robbins1952some, chernoff1959sequential}. 
Applying machine learning to automate hypothesis generation and testing goes back multiple decades \citep{king2004functional}. 
Optimal experiment design is amenable to Bayesian optimization \citep{greenhill2020bayesian} and reinforcement learning approaches \citep{kandasamy2019myopic}. While many OED approaches are unsuitable for our setting for the same reasons as BO and bandits, our method benefits from the application of Bayesian Algorithm Execution (BAX) \cite{neiswanger2021bayesian}, which we leverage as an acquisition function to identify candidate points with a high value of information. 

\section{Conclusion}
This work has presented a first step towards the development of optimal experiment design techniques targeted at the multi-stage drug discovery process. We have introduced a mathematical formalization of the drug discovery problem that captures the uncertainty inherent in the transition from in vitro to in vivo experiments. 
We proposed a novel algorithm based on Bayesian Algorithm Execution and illustrated its utility on many illustrative synthetic datasets. 
We have further evaluated this class of methods against the real-world large-scale assays from the GeneDisco benchmark, where they help identify diverse top interventions better than existing baselines.

A variety of exciting directions present themselves from the foundation laid by this paper. Future work could see the extension of the current framework to explicitly account for the fact that experimental cycles happen in batches, generalizing the iterative sampling approach considered here. 
Further, we assume in this work that distant representations of interventions imply different underlying biological mechanisms -- developing a causal formulation of the problem, and correspondingly identifying feature representations which capture this causal structure, would allow us to tell apart causally connected pathways more cleanly.
Finally, it is standard practice to measure several potential intermediate phenotypes of interest to capture different aspects of interest, which requires an extension of our approach to the setting of multiple objectives. 

%%%%%%%%%%%%%%%%%%%%%%%%

\section*{Acknowledgements}
CL was funded by an Open Philanthropy AI Fellowship when this work was completed. PN is supported by GSK and the UK Engineering and Physical Sciences Research Council (ESPRC ICASE award no.18000077). We thank the broader OATML and GSK.ai teams for helpful discussions throughout the project. Special thanks to Andreas Kirsch for his valuable feedback on the final manuscript and codebase.

\bibliography{refs}
\bibliographystyle{icml2023}

\newpage 

\appendix
\onecolumn
\input{appendix}

\end{document}

%% file: math_commands.tex
\usepackage{amsmath,amsfonts,bm, mathtools}
\def\eqref#1{equation~\ref{#1}}
\def\1{\bm{1}}

\def\rvx{{\mathbf{x}}}

\DeclareMathAlphabet{\mathsfit}{\encodingdefault}{\sfdefault}{m}{sl}
\SetMathAlphabet{\mathsfit}{bold}{\encodingdefault}{\sfdefault}{bx}{n}

\def\gA{{\mathcal{A}}}

\def\gD{{\mathcal{D}}}

\def\gG{{\mathcal{G}}}

\def\gP{{\mathcal{P}}}

\def\gX{{\mathcal{X}}}

\DeclareMathOperator*{\argmax}{arg\,max}
\DeclareMathOperator*{\argmin}{arg\,min}

\newcommand{\Gcal}{\mathcal{G}}
\newcommand{\Xcal}{\mathcal{X}}

\newcommand{\Pcal}{\mathcal{P}}
\newcommand{\RR}{\mathbb{R}}
\newcommand{\ip}{f_{\text{ip}}}
\newcommand{\iphat}{\widehat{f}_{\text{ip}}}

\newcommand{\sg}{S}
\newcommand{\dis}{f_{\text{out}}}
\newcommand{\dishat}{\widehat{f}_{\text{out}}}
\newcommand{\X}{\mathbf{X}}
\newcommand{\x}{\mathbf{x}}
\newcommand{\y}{\mathrm{y}}
\newcommand{\g}{\mathrm{g}}
\newcommand{\Out}{O}

\usepackage{amsthm}

\newtheorem{corollary}{Corollary}
\newtheorem{observation}{Observation}

%% file: genedisco_results.tex
\begin{table*}[t]
    \caption{\textbf{Performance comparison on GeneDisco CRISPR assays} We report the aggregated performance of \themethod{} and other methods on all assays from the GeneDisco benchmark. All other baselines and the breakdown per assay are provided in Appendix~\ref{appendix:genedisco_results}.}
    \label{table:genedisco_results}
    \begin{center}
    \begin{tabular*}{\textwidth}{ @{\extracolsep{\fill}} llccc}
    \toprule
    \textbf{Method} & \textbf{Category} & \textbf{Top-K recall} & \textbf{Diversity score} & \textbf{Overall score}\\
    \midrule 
Random & Random & 5.0\% (0.2\%) & 28.9\% (0.9\%) & 12.0\% (0.5\%) \\
Thompson Sampling & Bandits & 6.9\% (0.4\%) & 38.8\% (1.8\%) & 16.3\% (0.8\%) \\
UCB & Bayesian Optimization & 7.6\% (0.4\%) & 41.7\% (2.0\%) & 17.8\% (0.9\%) \\
\midrule 
Coreset & Active learning & 7.1\% (0.4\%) & 42.2\% (1.7\%) & 17.3\% (0.8\%) \\
JEPIG & Active learning & 7.3\% (0.4\%) & 41.5\% (1.8\%) & 17.4\% (0.9\%) \\
\midrule 
Levelset Bax & BAX & 7.9\% (0.4\%) & 43.1\% (1.9\%) & 18.5\% (0.9\%) \\
Top-K Bax & BAX & 7.8\% (0.4\%) & 44.6\% (1.9\%) & 18.7\% (0.9\%) \\
\midrule 
\textbf{DiscoBAX (ours)} & BAX & \textbf{8.1\% (0.5\%)} & \textbf{45.6\% (2.0\%)} & \textbf{19.2\% (1.0\%)} \\
\bottomrule
\end{tabular*}
\end{center}
\end{table*}

%% file: appendix.tex
\section{Biology background}
\label{sec:biology_background}

Here we provide the mathematical formalization of the engaged processes in the CRISPR-based gene knockout experiments from gene embeddings to assay readouts. We take a comprehensive approach for clarity but not all notations below are used in this work.

\begin{itemize}
    \item \textbf{Genes:} Let $\{g_1, g_2,\ldots ,g_m\}$ with $g_i\in\Gcal$ be all available genes for intervention.
    \item \textbf{Disease phenotype:} Several phenotype measurements are possible for every disease. Let $d\in D = \{d_1, d_2, \ldots, d_l\}$ be such a measurement from the list of $l$ possible readouts.
    \item \textbf{Intermediate phenotype functions:} Instead of the actual disease phenotype, intermediate readouts are used to measure the effect of a gene intervention on the disease phenotype. These readouts should be correlated with the downstream outcomes, but may present a simplified view of the disease action; for example, they might include the expression of certain proteins in a cancerous cell culture which are known to correlate with tumour growth rate (the disease phenotype). We let $ip\in IP =\{ip_1, ip_2, \ldots, ip_p|ip:D\to \RR\}$ be the set of maps from disease phenotype to real numbers that are the intermediate readouts for the effect of each gene intervention.
    \item \textbf{Knock-out function:} $\psi: G^m \to \Pcal(G)$ shows which genes to intervene on. It takes the set of all available genes as input and returns the subset of genes to get knocked out.
    \item \textbf{Disease mechanism function:} $f:G\times \Pcal(G) \to D^l$. This function takes all available genes and also the intervened subset and returns how the effect of the intervention on disease phenotype.
    \item \textbf{Knock-out representation} $\phi_{ko}: \Pcal(G)\to \RR^{d_{ko}}$ takes the subset of genes to knock out and returns a real-valued vector as the representation of this intervention.
    
    \item \textbf{Learnable mechanism:} To make the disease mechanism function amenable to learning algorithms, we use the intervention representation in the input and intermediate phenotype read-out in the output and work with $\{F_j: F_j=ip_j \circ f \circ \phi_{ko}^{-1} \textrm{for} 1\leq j \leq p\}$ where $F_j:\RR^d_{ko}\to \RR$ is the effect of a knock-out represented by the knock-out representation $\phi_{ko}$ in the input on the $j^{\textrm{th}}$ intermediate phenotype read-out in the output.
\end{itemize}

It is natural to work with real-valued functions with real-value domain which are more friendly to function estimation algorithms. For example, one can use MSE error as a metric to learn the intervention-to-assay mechanism from the available labeled datasets $D = \{(x_i, y_i)\}_{i=1}^n$ using the objective function
\begin{equation}
\label{eq:learnable_mechanism_loss}
    \hat{F}_j = \frac{1}{n}\sum_{(x_i, y_i)}\argmin_{F} \lVert F(x_i) -y_i)\rVert.
\end{equation}
for every $j$ that gives $\{\hat{F}_1, \hat{F}_2, \ldots, \hat{F}_p\}$. Notice that $\hat{y} = \hat{F}_j(x)$ is the best predictor of the intermediate disease phenotype (screen, assay) for the gene intervention $\psi(G^m)$ represented by $x=\phi_{ko}(\psi(G^m))$.

\section{Deeper insights into the DiscoBAX algorithm}
\label{apx:deeper-insights}
\subsection{Insights of \cref{eq:master_equation}: } 
\begin{figure}[t]

\centering
\includegraphics[width=\textwidth]{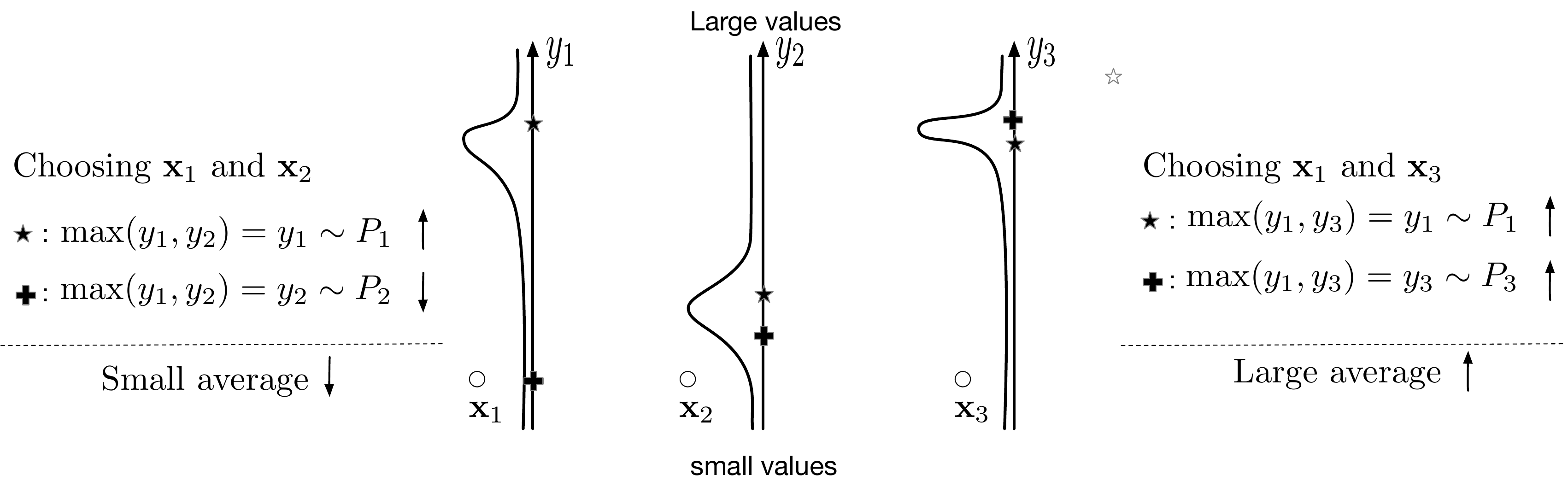}

\caption{A simple visualization to gain insight into \cref{eq:master_equation}. }
\label{fig:master_equation_intuition}
\end{figure}

In this section, we design a simplistic scenario to provide more insight into the proposed objective function \cref{eq:master_equation} and how it serves two purposes, i.e., choosing a set of interventions with high phenotype values and high diversity.
For convenience, in \cref{fig:master_equation_intuition}, we show a simple scenario where 2 out of 3 genes are to be chosen, i.e., $\lvert\mathcal{S}\rvert=2$ and $\lvert\mathcal{X}\rvert=3$. There are three ways of choosing a pair of genes out of three options. We aim to show which pair is favoured by \cref{eq:master_equation}. Without loss of generality, assume $\mathbf{x}_1$ is chosen. Due to the probabilistic model of $f_\text{out}$, all $y_i=f_\text{out}(x_i), i=1,2,3$ are random variables whose probability densities ($P_i$) are plotted next to each gene. It is observed that $P_1$ and $P_3$ are concentrated at larger values (higher regions of the vertical axis) compared to $P_2$ that puts much of its mass at lower values. Hence, in most realization, $y_1$ takes a large value (star) and $y_1 \approx \max(y_1, \cdot)$ as the second argument is sampled from distributions concentrated at lower values ($P_2$) or almost equally large values ($P_3$). The second argument becomes important in the rare events when $y_1$ takes a small value (cross). In this case, the output of $\max(y_1, \cdot)$ is no longer determined by its first argument and is, with high probability, influenced by the second argument which takes on a large value if realized from $P_3$ compared with $P_2$ as the former is concentrated at larger values. Hence, choosing the pair $(\mathbf{x}_1, \mathbf{x}_3)$ produces a larger average than $(\mathbf{x}_1, \mathbf{x}_2)$ and is therefore favourable by \cref{eq:master_equation}. Moreover, it is implicitly assumed in the above reasoning that $P_1$, $P_2$ and $P_3$ are not highly correlated. Otherwise, a small value of $y_1$ led to a small value of $y_3$ as well. Hence, \cref{eq:master_equation} chooses the genes which produce large values and the mechanisms that are modeled as the random effect are as independent as possible. This choice of genes increases the chance that if one gene fails to proceed to further steps of the drug discovery pipeline for some reason such as safety, tractability, etc, the other chosen genes will preserve high chances of success as they are likely to be involved in mechanisms different from those that cause the failure of the previous gene.

\subsection{Insights of \cref{fig:overview}: }This illustrates the motivation and goal of this research which is finding mathematical formulation and practical implementation of an algorithm that meets the actual needs of initial stages of drug discovery pipeline that neither value-seeking nor diversity-seeking methods can fulfill. The phenotypic effect of genetic perturbation can follow a complex function with many modes. We are mainly interested in genes which cause large changes in the measured phenotype as those are the genes that engage more with the disease and can be a potential target for a drug compound. However, as the figure shows, the value-seeking methods stop after finding one mode of the function (the light gray triangles which are concentrated in one of the modes but do not cover the other modes which have equally large values). This is risky since the genes that are associated with that mode are probably correlated in the sense that if one of them fails in the further steps of the drug discovery pipeline, the other may also fail with high likelihood. On the other end of the spectrum, although a diversity-seeking algorithm proposes uncorrelated genes that are unlikely to fail together (the dark gray circles which cover a large domain but miss the modes), it is highly inefficient and a large number of chosen genes may not be highly involved in the disease mechanism. Hence, the nature of the problem requires a middle-ground method that seeks the modes of the underlying function but covers as many does as possible (the red stars that efficiently cover all modes but not in-between spaces) so that if the genes associated with one mode fails, those associated with the other modes have chance to proceed in the pipeline.

\section{Sub-modularity of $S$}
\label{appendix:submodular}
\begin{observation}
The score function $S: \gP(\gG) \rightarrow \mathbb{R}$ defined by
\begin{equation}
    S(G) = \mathbb{E}_{\dis}[\max_{g \in G} \max (0, \dis(g)) ]
\end{equation}
is monotone submodular.
\end{observation}
\begin{proof}
%Straightforward: 
We first show monotonicity.
\begin{align*}
S(G \cup \{g\}) &= \mathbb{E}_{\dis}[\max_{g' \in G \cup \{g\}} \max (0, \dis(g')) ]\\
&= \mathbb{E}_{\eta}[\max_{g' \in G \cup \{g\}} \max (0, \dis(g', \eta)) ]\\
&\leq \mathbb{E}_{\eta}\max_{G} [\max (0, \dis(g', \eta)) + \max(0, \dis(g, \eta))] \\
&= \mathbb{E}_{\eta}\max_{G} [\max (0, \dis(g', \eta))] + \mathbb{E}_{\eta}[ \max(0, \dis(g, \eta)) ]\\
&= S(G) + S(\{g\})
\end{align*}
The proof for submodularity follows similarly. Letting $X \subseteq Y$ we have that $S$ is submodular if for any point $g$ we have $S(X \cup \{g\} ) - S(X) \geq S(Y \cup \{g\}) - S(Y)$. First, recall:
\begin{align}
S(X \cup \{x\}) - S(X) = \mathbb{E}_{\dis}[\max_{g' \in X \cup \{g\} }& \max (0, \dis(g'))] - \mathbb{E}_{\dis}[\max_{g' \in X} \max (0, \dis(g'))] \\
\intertext{We consider a single realization of the outcome $\dis$, and will show that the inequality holds for this outcome. If the maximum of $\dis$ over $Y$ is negative, then the result is trivial. Otherwise, there are three cases to consider: first, if $g$ maximizes $\dis$ over the set $Y \cup \{g\}$, then we have}
\max_{g' \in X \cup \{g\} } \max (0, \dis(g')) - \max_{g' \in X} \max (0, \dis(g')) &= \dis(g) - \max_{g' \in X} \max(0, \dis(g'))\\
&\geq \dis(g) - \max_{g' \in Y} \max(0, \dis(g'))\\
\intertext{as $X \subseteq Y$. Next, if $g$ does not maximize $\dis$ in $X$, then the difference on both sides of the inequality will be zero. Finally, if $g$ maximizes $\dis$ in $X$ but not in $Y$, we have the following:}
\max_{g' \in X \cup \{g\} } \max (0, \dis(g')) - \max_{g' \in X} \max (0, \dis(g')) &= \dis(g) - \max_{g' \in X} \max(0, \dis(g')) \\
& > 0 = \max_{g' \in Y \cup \{g\}} \max(0, \dis(g')) - \max_{g' \in Y} \max(0, \dis(g')) \\
\intertext{Since the inequality holds for each random realization of $\dis$, it applies to the expectation, and so we have}
 \mathbb{E}_{\eta}[\max_{g' \in X \cup \{g\} } \max (0, \dis(g'))] - \mathbb{E}_{\eta}[\max_{g' \in X} \max (0, \dis(g'))] &\geq  \mathbb{E}_{\eta}[\max_{g' \in Y \cup \{g\} } \max (0, \dis(g'))]\\ & - \mathbb{E}_{\eta}[\max_{g' \in Y} \max (0, \dis(g'))] 
\end{align}

\end{proof}

\begin{corollary}
The greedy algorithm which iteratively selects points maximizing $S(G)$ is a  $1- 1/e$ approximation of the optimal.
\end{corollary}

\section{Detailed experimental results}
\label{app:additional-results}
For reproducibility, the entire codebase for all experiments in this work can be found in the supplementary material attachment. Experiments with DiscoBAX in this section were conducted with Bernoulli noise. We present results with Gaussian noise in the next section (GeneDisco experiments).

\subsection{Synthetic dataset experiments}

\subsubsection{Sample complexity}

\label{apx:synthetic-experiments}
\begin{figure}
    \centering
    \includegraphics[width=0.32\linewidth]{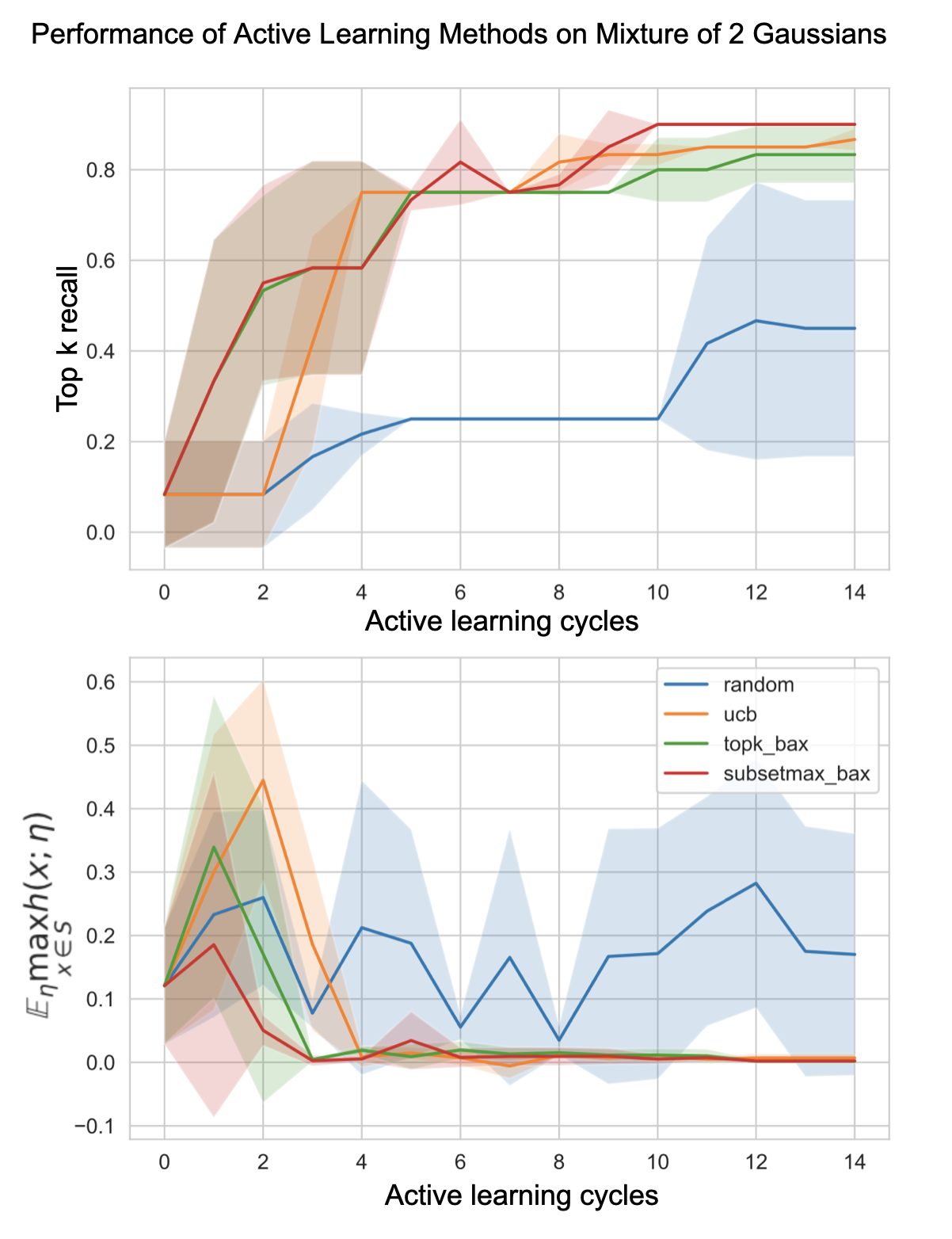}
    \includegraphics[width=0.32\linewidth]{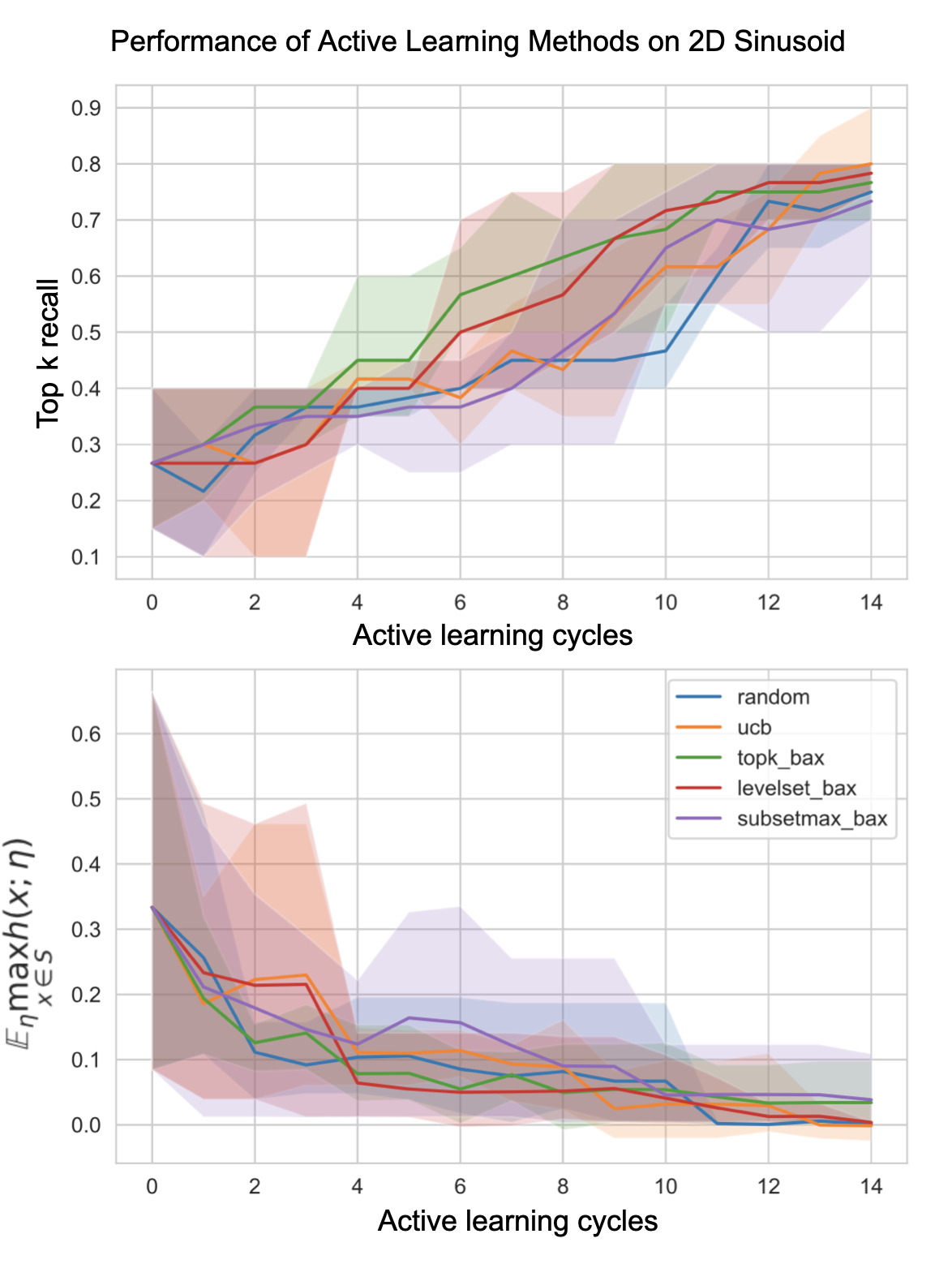}
    \includegraphics[width=0.32\linewidth]{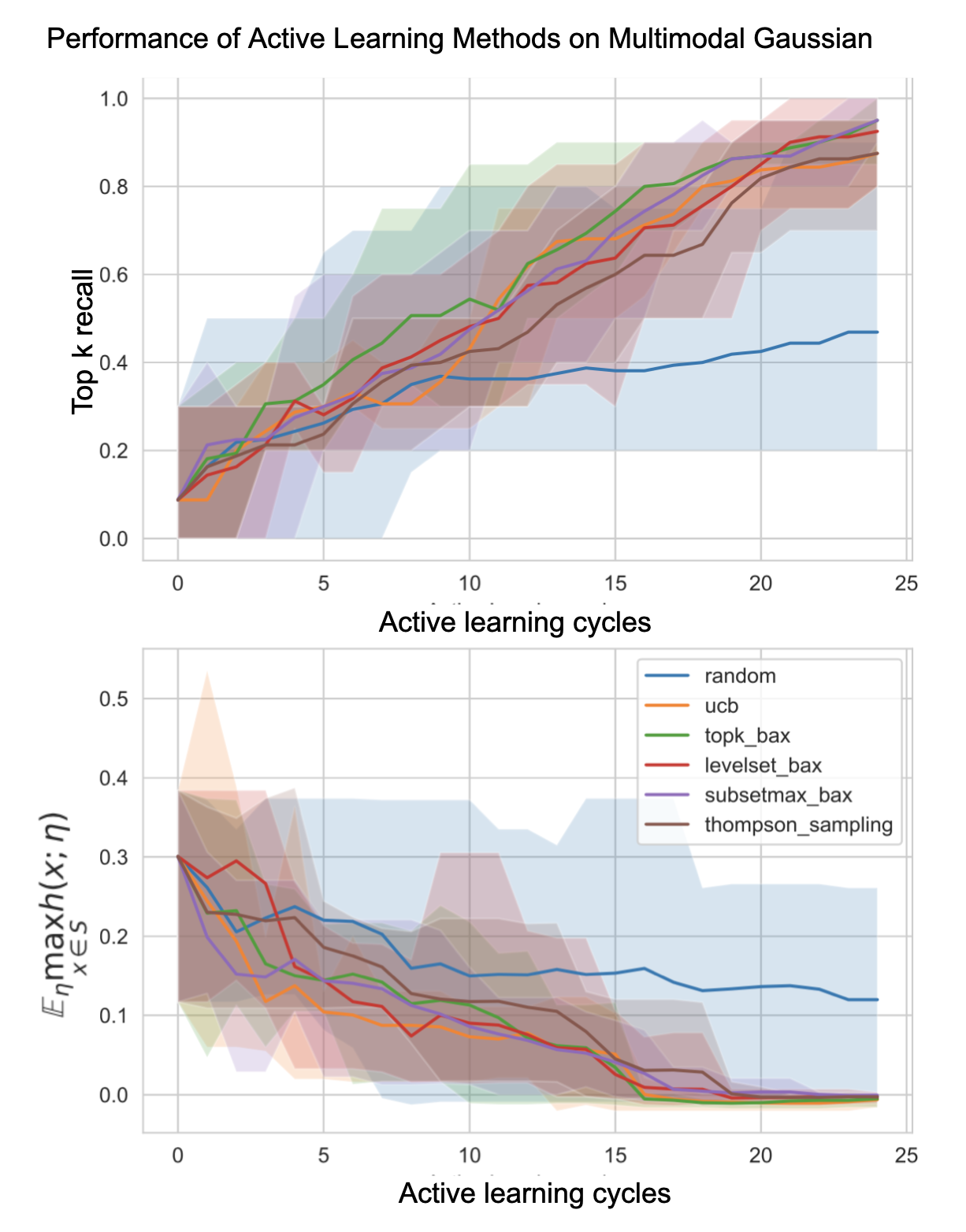}
    \caption{Top-k recall and expected maximal intervention value on: a) a mixture of two RBF kernels; b) a one-dimensional linear combination of sinusoids with multiple local optima; c) a mixture of four RBF kernels of varying scales.}
    \label{fig:synthetic-experiments_appendix}
\end{figure}
Our objective in this section is to validate a number of properties of the proposed method in interpretable synthetic datasets. 
\begin{itemize}
    \item \textbf{Sample complexity:} our method requires fewer samples to reach a global optimum relative to random sampling or naive uncertainty maximization methods.
    \item \textbf{Diversity of candidate set:} unlike standard Bayesian optimization methods, our approach identifies a set of points which approximately maximize the function while also maintaining diversity with respect to a pre-chosen metric, improving the robustness of the candidate set to uncertainty in the mapping between observable and terminal outcomes.
\end{itemize}
In these experiments and in Figure~\ref{fig:toy-example} we consider a number of baselines, including the following.

\begin{itemize}
    \item \textbf{Random:} $\x^* \sim \mathrm{Unif}(\gX \setminus \gD_t)$.
    \item \textbf{UCB:} naive upper-confidence sampling approach, letting $c \in \mathbb{R}$ be some constant: ${\x^* = \argmax_{\x \in \gX} \mu(\x) + c\sqrt{\sigma^2(\x) }.}$
    \item \textbf{BAX} acquisition (Algorithm~\ref{alg:active-learning}) for algorithm $\gA \in \{$Top-k, Levelset, Disco$\}$.
    \item \textbf{Thompson sampling:} acquisition based on maximum of sampled function from a Bayesian posterior. 
    ${\x^* = \argmax_{\x \in \gX} \iphat(\x) \quad \iphat \sim P(\ip|\gD_{\mathrm{train}})}.$
    \item \textbf{Active sampling:} maximize uncertainty over the input set ${
        \x^* = \argmax_{\x \in \gX} \sigma^2(\x)}.$
\end{itemize}

We consider the following synthetic datasets, where for all synthetic experiments we use a batch size equal to one.

\textbf{Mixture-of-Gaussians:} pdf of a mixture of gaussians with means [-0.5, 0.5], variances 0.1 and relative weights [0.3, 0.7]. $x \in [-1,1]$.

\textbf{Multimodal mixture:} given domain $[-7, 7]$, outputs the (scaled) density of a mixture of Gaussians with means $\{-4, -2, 0, 3\}$, variances $\{0.3, 0.35, 0.3,  0.35\}$, and weights $\{0.6, 0.45, 0.5, 0.4\}$.
\textbf{2-d sinusoid:}  $f(x) = \sin \bigg [  \frac{1}{2}\begin{pmatrix}
0.25 & -\frac{1}{\pi} \\
0.1 & .02
\end{pmatrix} \rvx \bigg ]$, $\rvx \in \mathbb{R}^2$, $-\pi < \rvx < \pi$

\subsubsection{Additional Empirical Evaluations on Synthetic Dataset}

\begin{figure}
    \centering
    \includegraphics[width=0.3\linewidth]{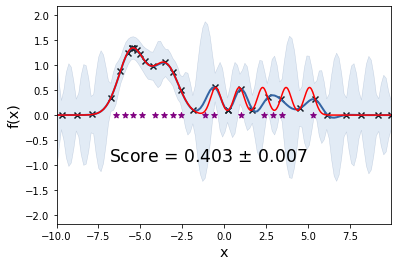}
    \caption{Evaluation of the EI acquisition function on the regression problem discussed previously.}
    \label{fig:expected-improvement}
\end{figure}

We include an evaluation of the Expected Improvement (EI) acquisition function (Fig.~\ref{fig:expected-improvement}) on the same task as was previously illustrated in Figure~\ref{fig:toy-example}. Because we use an acquisition batch size of one in these experiments, the parallel acquisition strategy qEI coincides with the incremental expected improvement acquisition function. Concretely, the expected improvement acquisition function performs the following maximization, given some pool $\mathcal{D}$ of already sampled points:
\begin{equation}
   \max_{x \not \in \mathcal{D}} \mathbb{E}_{P(f(x)|\mathcal{D})} \max( f(x) - f(x^*), 0)
\end{equation}
where $x^*$ is the element of $\mathcal{D}$ which maximizes $f$ and $P(\cdot | \mathcal{D})$ denotes the posterior over function values $f(x)$ for a fixed $x$.

\subsection{GeneDisco experiments}

\subsubsection{Clustering of optimal interventions}
\label{appendix:clustering}

In the GeneDisco experiments (\S~\ref{sec:genedisco}), we define a diversity metric based on the recall of Top-K clusters. These clusters are obtained for each assay as follows. All experiments we carried out in \S~\ref{sec:genedisco} leverage the Achilles dataset \citep{Dempster720243} from GeneDisco to represent the different interventions. This dataset characterizes each gene with an 808-dimensional vector. We first select the optimal interventions as the ones in the top percentile of disease phenotype for a given assay. We then project the Achilles representations of each intervention into a lower-dimensional subspace of dimension 20 with PCA. We then fit a Gaussian Mixture Model (GMM) with 20 mixtures to obtain the different clusters, selecting the best result out of 20 random initializations.

\subsubsection{Detailed performance analysis}
\label{appendix:genedisco_results}

We provide below detailed results across the five CRISPR assays from the GeneDisco benchmark: the Interferon $\gamma$ and Interleukin 2 assays based on \cite{Schmidt2021Cytokine}, the Leukemia assay with NK cells from \cite{zhuang2019LeukemiaNK}, the SARS-CoV-2 assay from \cite{zhu2021genome} and the Tau protein assay from \cite{sanchez2021genome}. All interventions for the five assays were represented based on the Achilles dataset \citep{Dempster720243}. For the active learning baselines already present in GeneDisco we used the same hyperparameters as in \cite{mehrjou2021genedisco}. For the additional baselines introduced in this work, we use standard/default hyperparameters everywhere (see our codebase for all details), except for the dedicated hyperparameter analysis in Appendix~\ref{appendix:genedisco_hypers}. 
We used DiscoBAX with Gaussian noise (with length scale for the underlying Radial Basis Function kernel equal to 1) in the results below, but obtain comparable performance with Bernoulli noise. To prevent model overfitting during the various acquisition cycles, we closely followed experimental protocol in \citet{mehrjou2021genedisco} and selected similar model architectures and hyperparameters.

\begin{figure}
    \centering
    \includegraphics[width=0.85\linewidth]{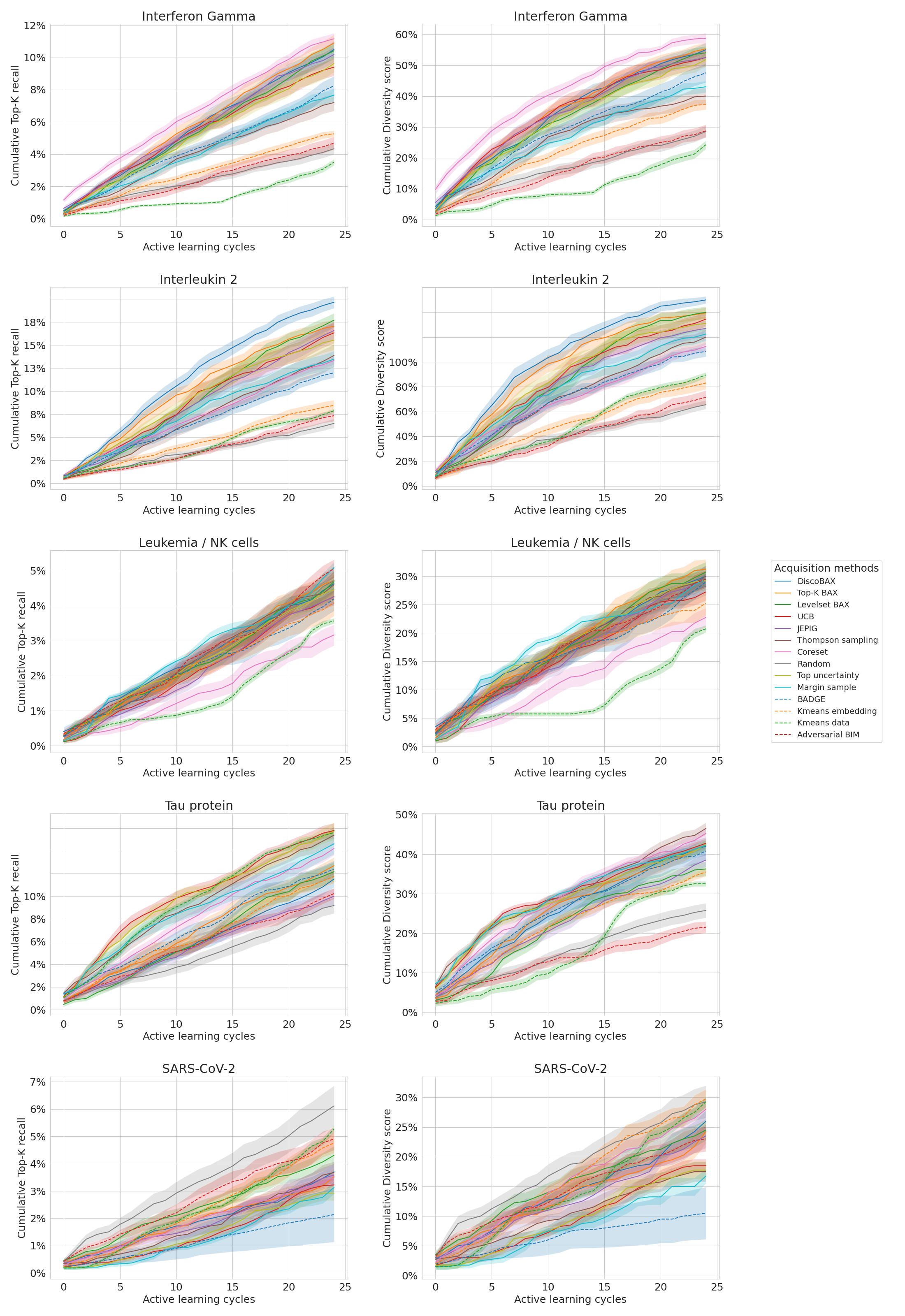}
    \caption{\textbf{Top-K recall and Diversity score Vs acquisition cycles for all GeneDisco CRISPR assays}}
    \label{fig:genedisco_main_results_appendix}
\end{figure}

We observe in 
\Cref{table:genedisco_results,table:genedisco_detailed_results_interleukin,table:genedisco_detailed_results_sars,table:genedisco_detailed_results_leukemia,table:genedisco_detailed_results_tau} that DiscoBAX outperforms all other baselines we compare against in aggregate. The superior sample efficiency of the scheme is apparent Fig.~\ref{fig:genedisco_main_results} and Fig.~\ref{fig:genedisco_main_results_appendix} as DiscoBAX exhibits high recall and diversity score throughout the different learning cycles.
Across the different assays, DiscoBAX and other BAX methods tend to perform consistently high, while other approaches such as Coreset or UCB, achieve high performance on 1 or 2 assays, but do poorly elsewhere. 
As discussed in section \ref{sec:genedisco} and as noted in \citet{mehrjou2021genedisco}, the fact that ‘Random’ outperforms all other baselines on that dataset seems to indicate an issue with the data (eg., the feature space does not correlate with the disease phenotype, high label noise) rather than an algorithmic issue (‘Random’ performs very poorly on all other 4 assays). 

\input{genedisco_detailed_tables.tex}

\subsection{GeneDisco experiments - hyperparameter selection}
\label{appendix:genedisco_hypers}

For the three BAX algorithms (Top-K BAX, Levelset BAX and DiscoBAX), we optimize the main hyperparameters of each method (respectively the K parameter, the level threshold and the number S of sets in SubsetSelect). To mitigate the risk of overfitting, we select our hyperparameters based on a single assay (the `Tau protein' assay), and use the obtained optimal values in experiments for all assays. We perform a grid search for each hyperparameter, repeating each experiment over 20 seeds. We find that, on that dataset, optimal values for the hyperparameters are respectively K=2, level=1.5 and S=10. Importantly, we find that the performance of DiscoBAX is substantially more robust to the choice of hyperparameters compared with the other two BAX algorithms (Table~\ref{appendix:genedisco_hypers}).

\input{genedisco_hyper_table.tex}

%% file: genedisco_detailed_tables.tex
    \begin{table}[!ht]
    \caption{\textbf{Detailed performance comparison on GeneDisco - Interferon $\gamma$ assay}. The top 3 models are bolded.}
    \label{table:genedisco_detailed_results_interferon}
    \begin{center}
    %\begin{small}
    %\resizebox{\textwidth}{!}{
    %@{\extracolsep{\fill}}
    \begin{tabular}{llccc}
    \toprule
    \textbf{Dataset} & \textbf{Method} & \textbf{Top-K recall} & \textbf{Diversity score} & \textbf{Overall score}\\
    \midrule 
     \multirow{14}{*}{Interferon $\gamma$} & Adversarial BIM & 4.7\% (0.4\%) & 28.8\% (2.1\%) & 11.6\% (0.9\%) \\
& Badge & 8.2\% (0.6\%) & 47.5\% (3.2\%) & 19.8\% (1.4\%) \\
& Coreset & \textbf{11.2\% (0.4\%)} & \textbf{58.8\% (1.7\%)} & \textbf{25.6\% (0.8\%)} \\
& DiscoBAX (ours) & \textbf{10.5\% (0.5\%)} & \textbf{55.0\% (2.4\%)} & \textbf{24.0\% (1.1\%)} \\
& JEPIG & 10.2\% (0.5\%) & 52.5\% (2.3\%) & 23.1\% (1.1\%) \\
& Kmeans Data & 3.5\% (0.2\%) & 24.3\% (1.4\%) & 9.2\% (0.5\%) \\
& Kmeans Embedding & 5.3\% (0.2\%) & 37.3\% (1.7\%) & 14.0\% (0.6\%) \\
& Levelset Bax & 10.4\% (0.5\%) & 54.0\% (2.3\%) & 23.7\% (1.1\%) \\
& Marginsample & 7.7\% (0.4\%) & 43.0\% (1.9\%) & 18.1\% (0.9\%) \\
& Random & 4.3\% (0.4\%) & 28.5\% (2.0\%) & 11.1\% (0.8\%) \\
& Soft Uncertainty & 5.2\% (0.6\%) & 32.0\% (3.2\%) & 12.9\% (1.3\%) \\
& Thompson Sampling & 7.2\% (0.5\%) & 40.0\% (2.7\%) & 17.0\% (1.2\%) \\
& Top-K Bax & \textbf{10.9\% (0.5\%)} & \textbf{55.3\% (2.1\%)} & \textbf{24.5\% (1.0\%)} \\
& Top Uncertainty & 9.6\% (0.6\%) & 51.8\% (2.3\%) & 22.3\% (1.1\%) \\
& UCB & 9.4\% (0.6\%) & 52.5\% (2.9\%) & 22.2\% (1.3\%) \\

    \bottomrule
    \vspace{-0.8cm}
    \end{tabular}
    %}
    %\end{small}
    \end{center}
    \end{table}

    \begin{table}[!ht]
    \caption{\textbf{Detailed performance comparison on GeneDisco - Interleukin 2 assay}. The top 3 models are bolded.}
    \label{table:genedisco_detailed_results_interleukin}
    \begin{center}
    %\begin{small}
    %\resizebox{\textwidth}{!}{
    %@{\extracolsep{\fill}}
    \begin{tabular}{llccc}
    \toprule
    \textbf{Dataset} & \textbf{Method} & \textbf{Top-K recall} & \textbf{Diversity score} & \textbf{Overall score}\\
    \midrule 
     \multirow{14}{*}{Interleukin 2} & Adversarial BIM & 5.9\% (0.5\%) & 35.8\% (2.9\%) & 14.5\% (1.2\%) \\
& Badge & 9.6\% (0.4\%) & 54.3\% (2.2\%) & 22.8\% (1.0\%) \\
& Coreset & 10.7\% (0.4\%) & 56.3\% (1.8\%) & 24.5\% (0.8\%) \\
& DiscoBAX (ours) & \textbf{15.7\% (0.5\%)} & \textbf{75.0\% (1.5\%)} & \textbf{34.3\% (0.9\%)} \\
& JEPIG & 13.3\% (0.8\%) & 63.5\% (3.4\%) & 29.0\% (1.7\%) \\
& Kmeans Data & 6.3\% (0.2\%) & 44.8\% (1.1\%) & 16.8\% (0.5\%) \\
& Kmeans Embedding & 6.8\% (0.5\%) & 41.5\% (2.6\%) & 16.8\% (1.1\%) \\
& Levelset Bax & \textbf{14.2\% (0.6\%)} & \textbf{70.0\% (2.4\%)} & \textbf{31.5\% (1.2\%)} \\
& Marginsample & 10.8\% (0.7\%) & 61.3\% (2.8\%) & 25.7\% (1.4\%) \\
& Random & 5.2\% (0.4\%) & 32.8\% (1.8\%) & 13.1\% (0.9\%) \\
& Soft Uncertainty & 4.9\% (0.6\%) & 29.0\% (3.0\%) & 11.9\% (1.3\%) \\
& Thompson Sampling & 11.1\% (1.0\%) & 60.0\% (3.2\%) & 25.8\% (1.8\%) \\
& Top-K Bax & \textbf{13.6\% (0.6\%)} & \textbf{69.8\% (2.3\%)} & \textbf{30.8\% (1.2\%)} \\
& Top Uncertainty & 12.4\% (0.9\%) & 65.5\% (3.0\%) & 28.5\% (1.6\%) \\
& UCB & 13.1\% (0.9\%) & 67.3\% (3.0\%) & 29.6\% (1.7\%) \\
    
    \bottomrule
    \vspace{-0.8cm}
    \end{tabular}
    %}
    %\end{small}
    \end{center}
    \end{table}

    \begin{table}[!ht]
    \caption{\textbf{Detailed performance comparison on GeneDisco - SARS-CoV-2 assay}. The top 3 models are bolded.}
    \label{table:genedisco_detailed_results_sars}
    \begin{center}
    %\begin{small}
    %\resizebox{\textwidth}{!}{
    %@{\extracolsep{\fill}}
    \begin{tabular}{llccc}
    \toprule
    \textbf{Dataset} & \textbf{Method} & \textbf{Top-K recall} & \textbf{Diversity score} & \textbf{Overall score}\\
    \midrule 
     \multirow{14}{*}{SARS-CoV-2} & Adversarial BIM & 4.9\% (0.4\%) & 23.0\% (2.2\%) & 10.6\% (1.0\%) \\
& Badge & 2.1\% (1.0\%) & 10.5\% (4.5\%) & 4.7\% (2.2\%) \\
& Coreset & 3.5\% (0.2\%) & 28.0\% (1.9\%) & 9.9\% (0.7\%) \\
& DiscoBAX (ours) & 3.7\% (0.3\%) & 26.0\% (1.9\%) & 9.8\% (0.7\%) \\
& JEPIG & 3.6\% (0.4\%) & 23.5\% (2.1\%) & 9.2\% (0.9\%) \\
& Kmeans Data & \textbf{5.3\% (0.1\%)} & \textbf{29.3\% (0.9\%)} & \textbf{12.4\% (0.4\%)} \\
& Kmeans Embedding & 4.8\% (0.4\%) & \textbf{29.8\% (1.7\%)} & \textbf{11.9\% (0.8\%)} \\
& Levelset Bax & 4.3\% (0.4\%) & 24.5\% (1.8\%) & 10.3\% (0.8\%) \\
& Marginsample & 3.1\% (0.2\%) & 16.8\% (1.3\%) & 7.2\% (0.6\%) \\
& Random & \textbf{6.1\% (0.8\%)} & \textbf{29.3\% (2.8\%)} & \textbf{13.4\% (1.5\%)} \\
& Soft Uncertainty & \textbf{7.3\% (5.0\%)} & 14.5\% (6.0\%) & 10.3\% (5.5\%) \\
& Thompson Sampling & 3.7\% (0.4\%) & 17.5\% (1.7\%) & 8.0\% (0.8\%) \\
& Top-K Bax & 3.5\% (0.3\%) & 24.3\% (2.0\%) & 9.2\% (0.8\%) \\
& Top Uncertainty & 2.9\% (0.3\%) & 17.8\% (1.6\%) & 7.2\% (0.7\%) \\
& UCB & 3.2\% (0.2\%) & 18.5\% (1.2\%) & 7.7\% (0.5\%) \\

    \bottomrule
    \vspace{-0.8cm}
    \end{tabular}
    %}
    %\end{small}
    \end{center}
    \end{table}

    \begin{table}[!ht]
    \caption{\textbf{Detailed performance comparison on GeneDisco - Leukemia/NK assay}. The top 3 models are bolded.}
    \label{table:genedisco_detailed_results_leukemia}
    \begin{center}
    %\begin{small}
    %\resizebox{\textwidth}{!}{
    %@{\extracolsep{\fill}}
    \begin{tabular}{llccc}
    \toprule
    \textbf{Dataset} & \textbf{Method} & \textbf{Top-K recall} & \textbf{Diversity score} & \textbf{Overall score}\\
    \midrule 
     \multirow{14}{*}{Leukemia/NK} & Adversarial BIM & \textbf{5.1\% (0.3\%)} & 29.5\% (1.6\%) & \textbf{12.2\% (0.7\%)} \\
& Badge & 4.2\% (0.4\%) & 29.0\% (1.6\%) & 11.0\% (0.8\%) \\
& Coreset & 3.2\% (0.3\%) & 22.8\% (1.7\%) & 8.5\% (0.7\%) \\
& DiscoBAX (ours) & 4.6\% (0.2\%) & \textbf{30.0\% (1.9\%)} & 11.8\% (0.7\%) \\
& JEPIG & 4.3\% (0.2\%) & 29.5\% (1.4\%) & 11.2\% (0.5\%) \\
& Kmeans Data & 3.6\% (0.1\%) & 20.8\% (0.7\%) & 8.6\% (0.2\%) \\
& Kmeans Embedding & 4.1\% (0.4\%) & 25.3\% (2.2\%) & 10.2\% (0.9\%) \\
& Levelset Bax & 4.7\% (0.3\%) & \textbf{30.8\% (1.9\%)} & 12.0\% (0.7\%) \\
& Marginsample & \textbf{5.1\% (0.2\%)} & 29.5\% (1.4\%) & \textbf{12.2\% (0.5\%)} \\
& Random & \textbf{4.7\% (0.4\%)} & 28.3\% (1.7\%) & 11.5\% (0.8\%) \\
& Soft Uncertainty & 4.7\% (0.4\%) & 28.8\% (2.1\%) & 11.6\% (0.9\%) \\
& Thompson Sampling & 4.6\% (0.4\%) & 30.0\% (2.0\%) & 11.7\% (0.9\%) \\
& Top-K Bax & 4.7\% (0.3\%) & \textbf{31.3\% (1.8\%)} & \textbf{12.1\% (0.7\%)} \\
& Top Uncertainty & 4.5\% (0.3\%) & 29.3\% (1.8\%) & 11.4\% (0.8\%) \\
& UCB & 4.3\% (0.2\%) & 27.3\% (1.6\%) & 10.8\% (0.6\%) \\

    \bottomrule
    \vspace{-0.8cm}
    \end{tabular}
    %}
    %\end{small}
    \end{center}
    \end{table}
    
    \begin{table}[!ht]
    \begin{center}
    \caption{\textbf{Detailed performance comparison on GeneDisco - Tau protein assay}. The top 3 models are bolded.}
    \label{table:genedisco_detailed_results_tau}
    %\begin{small}
    %\resizebox{\textwidth}{!}{
    \begin{tabular}{llccc}
    \toprule
    \textbf{Dataset} & \textbf{Method} & \textbf{Top-K recall} & \textbf{Diversity score} & \textbf{Overall score}\\
    \midrule 
     \multirow{14}{*}{Tau protein} & Adversarial BIM & 5.1\% (0.2\%) & 21.5\% (1.5\%) & 10.5\% (0.6\%) \\
& Badge & 6.2\% (0.4\%) & 40.8\% (2.3\%) & 15.9\% (0.9\%) \\
& Coreset & 7.1\% (0.3\%) & \textbf{45.3\% (1.6\%)} & 17.9\% (0.7\%) \\
& DiscoBAX (ours) & 5.8\% (0.4\%) & 42.0\% (2.4\%) & 15.6\% (0.9\%) \\
& JEPIG & 5.0\% (0.4\%) & 38.5\% (1.7\%) & 13.9\% (0.8\%) \\
& Kmeans Data & \textbf{7.8\% (0.1\%)} & 32.5\% (0.7\%) & 15.9\% (0.3\%) \\
& Kmeans Embedding & 5.9\% (0.3\%) & 35.5\% (1.1\%) & 14.4\% (0.6\%) \\
& Levelset Bax & 6.1\% (0.4\%) & 36.3\% (1.8\%) & 14.8\% (0.9\%) \\
& Marginsample & 7.3\% (0.5\%) & 42.3\% (1.5\%) & 17.6\% (0.8\%) \\
& Random & 4.6\% (0.4\%) & 25.8\% (1.9\%) & 10.9\% (0.8\%) \\
& Soft Uncertainty & 4.6\% (0.4\%) & 29.0\% (1.7\%) & 11.6\% (0.8\%) \\
& Thompson Sampling & 7.7\% (0.4\%) & \textbf{46.5\% (1.5\%)} & \textbf{18.9\% (0.8\%)} \\
& Top-K Bax & 6.3\% (0.4\%) & 42.3\% (1.6\%) & 16.4\% (0.8\%) \\
& Top Uncertainty & \textbf{7.9\% (0.4\%)} & 42.5\% (1.2\%) & \textbf{18.3\% (0.7\%)} \\
& UCB & \textbf{7.9\% (0.3\%)} & \textbf{42.8\% (1.2\%)} & \textbf{18.4\% (0.7\%)} \\

    \bottomrule
    \vspace{-0.8cm}
    \end{tabular}
    %}
    %\end{small}
    \end{center}
    \end{table}

%% file: genedisco_hyper_table.tex
\begin{table}[!ht]
    \caption{\textbf{GeneDisco experiment - Hyperparameter selection}}
    \label{table:genedisco_hyper_results}
    \begin{center}
    \begin{tabular}{lcccc}
    \toprule
    \textbf{Method} & \textbf{Hyperparameter value} & \textbf{Top-K recall} & \textbf{Diversity score} & \textbf{Overall score}\\
    \midrule 
     \multirow{6}{*}{Top-K BAX}    & 2 & 6.3\% (4.9\%) & 42.3\% (21.2\%) & \textbf{16.4\% (10.1\%)} \\
 & 3 & 5.4\% (4.7\%) & 39.0\% (24.2\%) & 14.6\% (10.6\%) \\
 & 5 & 5.0\% (6.0\%) & 37.5\% (37.1\%) & 13.7\% (15.0\%) \\
 & 10 & 5.0\% (4.5\%) & 32.5\% (32.0\%) & 12.7\% (12.0\%) \\
 & 20 & 5.0\% (4.2\%) & 26.8\% (33.6\%) & 11.6\% (11.8\%) \\
 & 32 & 5.1\% (3.7\%) & 28.0\% (34.9\%) & 11.9\% (11.4\%) \\
    \midrule
     \multirow{6}{*}{Levelset BAX}    & 0.5 & 4.3\% (4.9\%) & 31.3\% (29.6\%) & 11.6\% (12.1\%) \\
 & 0.8 & 4.3\% (3.4\%) & 28.5\% (16.3\%) & 11.1\% (7.4\%) \\
 & 1 & 4.6\% (4.5\%) & 31.5\% (19.4\%) & 12.1\% (9.3\%) \\
 & 1.1 & 4.8\% (4.4\%) & 30.3\% (27.1\%) & 12.0\% (10.9\%) \\
 & 1.2 & 5.4\% (3.6\%) & 32.0\% (27.1\%) & 13.1\% (9.8\%) \\
 & 1.5 & 6.1\% (5.7\%) & 36.3\% (23.9\%) & \textbf{14.8\% (11.7\%)} \\
    \midrule
     \multirow{6}{*}{DiscoBAX}   & 2 & 5.4\% (3.9\%) & 43.8\% (30.3\%) & 15.4\% (10.8\%) \\
 & 3 & 6.1\% (3.3\%) & 38.8\% (26.5\%) & 15.4\% (9.4\%) \\
 & 5 & 5.5\% (5.1\%) & 39.0\% (25.9\%) & 14.6\% (11.5\%) \\
 & 10 & 5.8\% (4.8\%) & 42.0\% (30.5\%) & \textbf{15.6\% (12.1\%)} \\
 & 20 & 5.0\% (5.2\%) & 40.3\% (25.5\%) & 14.1\% (11.5\%) \\
 & 32 & 5.3\% (4.7\%) & 38.3\% (27.8\%) & 14.3\% (11.5\%) \\
    \bottomrule
    \vspace{-0.8cm}
    \end{tabular}
    %}
    %\end{small}
    \end{center}
    \end{table}